\documentclass[preprint,aps,floats,nofootinbib,amssymb]{revtex4}
\usepackage{epsf,epsfig}

\usepackage{graphicx}

\newcommand{\be}{\begin{equation}}
\newcommand{\ee}{\end{equation}}
\newcommand{\bea}{\begin{eqnarray}}
\newcommand{\eea}{\end{eqnarray}}
\newcommand{\nn}{\nonumber}

\renewcommand{\check}{({\bf This statement needs to be checked!!})}

\newcommand{\mueff}{\mu_{\rm eff}}

\newcommand{\ba}{\begin{array}}
\newcommand{\ea}{\end{array}}
\newcommand{\bi}{\begin{itemize}}
\newcommand{\ei}{\end{itemize}}

\newcommand{\N}{\chi^0}
\newcommand{\C}{\chi^\pm}

\preprint{
\hbox to \hsize{
\hfill$\vcenter{\hbox{\bf MADPH-07-1476}
                \hbox{\bf hep-ph/0702036}
                \hbox{February 2007}}$}
}

\begin{document}

%-----------------------------------
% Title
%-----------------------------------
\title{\vspace*{.75in}
Recoil Detection of the Lightest Neutralino in MSSM Singlet Extensions}

%-----------------------------------
% Authors
%-----------------------------------
\author{
Vernon Barger$^1$,
Paul Langacker$^2$,
Ian Lewis$^1$,
Mat McCaskey$^1$,
Gabe Shaughnessy$^1$,
Brian Yencho$^1$
}

%-----------------------------------
% Address
%-----------------------------------
\affiliation{
$^1$Department of Physics, University of Wisconsin, Madison, WI 53706 \\
$^2$School of Natural Sciences, Institute for Advanced Study, Einstein Drive Princeton, NJ 08540\\
\vspace*{.5in}}

\thispagestyle{empty}

\begin{abstract}
\noindent
We investigate the correlated predictions of singlet extended MSSM models for direct detection of the lightest neutralino with its cosmological relic density.  To illustrate the general effects of the singlet, we take heavy sleptons and squarks.  We apply LEP, $(g-2)_\mu$ and perturbativity constraints.  We find that the WMAP upper bound on the cold dark matter density limits much of the parameter space to regions where the lightest neutralino can be discovered in recoil experiments.  The results for the NMSSM and UMSSM are typically similar to the MSSM since their light neutralinos have similar compositions and masses.  In the n/sMSSM the neutralino is often very light and its recoil detection is within the reach of the CDMS II experiment.  In general, most points in the parameter spaces of the singlet models we consider are accessible to the WARP experiment.
\end{abstract}
\maketitle
%=======================================================================
% BEGIN MAIN TEXT
%=======================================================================
%\newpage

%%%%%%%%%%%%%%%%%%%%%%%%%%%%
\section{Introduction}
%%%%%%%%%%%%%%%%%%%%%%%%%%%%
The evidence for cold, non-baryonic dark matter is one of the strongest cases for physics beyond the Standard Model (SM).  The most plausible candidates are axions and weakly interacting massive particles (WIMPS).  Many models have been proposed that provide a WIMP candidate for dark matter (DM).  These models include Supersymmetry \cite{Dimopoulos:1981zb,Baer:2004rs,Drees:2004jm}, Extra Dimensions \cite{Arkani-Hamed:1998rs,Randall:1999ee}, Little Higgs \cite{Arkani-Hamed:2001nc,Arkani-Hamed:2002qy,Schmaltz:2005ky}, and twin Higgs models \cite{Chacko:2005pe,Barger:1981dn}.  It is anticipated that the LHC and ILC may directly produce the DM particle, providing crucial information on its existence, origin and properties \cite{Baer:2003ru,Baltz:2006fm,Hooper:2006wv,Carena:2006nv,Feng:2005gj,Arnowitt:2007pi,Birkedal:2004xn,Cirigliano:2006dg}.  Recoils of nuclei from the scattering of WIMPS can also provide direct evidence for this DM candidate \cite{DMSAG,Vergados:2006ry}.  Alternative avenues exist to discover the DM particle indirectly.  These include observation of gamma ray lines in the galactic halo peaked at the neutralino mass \cite{Bergstrom:2001jj,Bergstrom:1997fj}, observation of high energy neutrinos from WIMP annihilation in the sun \cite{Barger:2001ur,Cirelli:2005gh,Profumo:2006bx}, and antiparticle detection \cite{Silk:1984zy,Baltz:2001ir,Hailey:2005yx,Mori:2001dv}.

The Minimal Supersymmetric Standard Model (MSSM) with conserved $R$-parity has a viable dark matter candidate, the lightest neutralino, that can naturally explain the abundance of cold dark matter (CDM) in the universe.  Other motivations for the MSSM include solutions to the gauge hierarchy problem, the quadratic divergence in the Higgs boson mass, and gauge coupling unification~\cite{Langacker:1991an,Ellis:1990wk,Amaldi:1991cn}.  The MSSM lagrangian contains a Higgsino mixing parameter, $\mu$,  the only massive parameter of the model that conserves supersymmetry.  The $\mu$ parameter must be at the electroweak (EW) or TeV scale to explain electroweak symmetry breaking.  This is the so-called hierarchy problem of supersymmetry because a priori the value of $\mu$ is arbitrary \cite{muproblem}.  The problem may be resolved by promoting the $\mu$ parameter to a dynamical field whose vacuum expectation value $\langle S\rangle$ and coupling $\lambda$ determine the effective $\mu$-parameter,
\be
\mu_{\rm eff} = \lambda \langle S\rangle.
\ee
Singlet extended models (xMSSM) have significant consequences for the Higgs and neutralino sectors \cite{Accomando:2006ga,Han:2004yd,Barger:2006kt,Barger:2006dh,Barger:2005hb,Choi:2004zx,Choi:2006fz,Gunion:2005rw,Ellwanger:2004gz,Arhrib:2006sx,Fayet:1974pd, Fayet:1977yc}.

In this paper we focus on the some of the proposed singlet extensions of the MSSM: (i) the Next-to-Minimal Supersymmetric SM (NMSSM) \cite{Ellis:1988er,NMSSM1,NMSSM2} with a trilinear singlet term in the superpotential, (ii) the Nearly-Minimal Supersymmetric SM (nMSSM) \cite{nMSSM, Panagiotakopoulos:2000wp,Menon:2004wv,Dedes:2000jp} with a tadpole term in the superpotential, and (iii) the $U(1)'$-extended MSSM (UMSSM) \cite{UMSSM1,umssm2,deCarlos:1997yv} with an extra $Z'$ gauge boson, as detailed in Table \ref{tbl:models} along with the respective symmetries and numbers of neutralino, chargino and Higgs states.  A Secluded $U(1)'$-extended MSSM (sMSSM) \cite{smssm,Han:2004yd} with three singlets in addition to the standard UMSSM Higgs singlet is equivalent to the nMSSM in the limit that the additional singlet vevs are large, and the trilinear singlet coupling, $\lambda_s$, is small \cite{Barger:2006dh}.   Thus, we collectively refer to the nMSSM and sMSSM as the n/sMSSM.  

Other singlet extensions are possible, and often are derived from string constructions.  The UMSSM is an example of such string constructions \cite{umssm2,Li:2006xb,King:2005jy,King:2005my,King:2007uj}.  For recent reviews and articles, see, e.g., Refs \cite{Blumenhagen:2006ci,Blumenhagen:2006ux,Blumenhagen:2005mu,Faraggi:2006qa,Li:2006xb,King:2005jy,King:2005my,King:2007uj}.  For a review of supersymmetric singlet models, see Ref. \cite{Accomando:2006ga}.  The SM may also be extended to include a singlet.  Depending on the parity of the singlet, it may mix with the Higgs boson \cite{O'Connell:2006wi,Bahat-Treidel:2006kx} or provide a viable dark matter candidate \cite{Burgess:2000yq,He:2007tt}.

\begin{table}[ht]
\begin{center}
\begin{tabular}{|c|clclclclc|}
\hline
Model:& MSSM &NMSSM &nMSSM& UMSSM & sMSSM\\
\hline
Symmetry:  &--  &~~ $\mathbb Z_3$    & $\mathbb Z^R_5, \mathbb Z^R_7$      & ~~$U(1)'$&$U(1)'$ \\\hline
Extra   &--         &       ~~${\kappa\over3} \hat S^3$    & $t_F  \hat S$& ~~-- &$\lambda_S \hat S_1 \hat S_2 \hat S_3$ \\
superpotential term&   --      &      (cubic)    & (tadpole) & ~~-- & (trilinear secluded)\\\hline
 $\N_i$ &        	4	  &	~~~5	&  	5	 & ~~6 & 9\\
 $H^0_i$ &       	2	  &	~~~3	&  	3	 & ~~3 & 6\\
 $A^0_i$ &        	1	  &	~~~2	&  	2	 & ~~1 & 4\\
%  $\C_i$ &        	2	  &	~~~2	&  	2	 & ~~2 & 2\\
% $H^{\pm}_i$ & 1	  &	~~~1	&  	1	 & ~~1 & 1\\
\hline
\end{tabular}
\caption{Symmetries associated with each model and their respective terms in the superpotential; the number of states in the neutralino and Higgs sectors are also given.  All models have two charginos, $\C_i$, and one charged Higgs boson, $H^\pm$.}
\label{tbl:models}
\end{center}
\end{table}

The extended MSSM models (xMSSM) have extra neutralino and Higgs states, that affect the spectra and composition of the particle states \cite{Barger:2005hb,Barger:2006kt}, and significantly influence the parameter space for the observation via direct or indirect detection of neutralino dark matter.  For example, the processes involved in indirect detection have been shown to be radiatively enhanced by the extended Higgs sector in the NMSSM \cite{Ferrer:2006hy}.  The direct detection prospects have been recently investigated in the NMSSM \cite{Cerdeno:2004xw,Cerdeno:2007sn}.  The relic density of the lightest neutralino can also be dependent on the model; the density relative to the closure density is very roughly given by the total annihilation cross-section by \cite{Yao:2006px}
\be
\Omega_{\N_1}h^2 \simeq \frac{0.1\text{ pb}}{\langle\sigma_{\rm ann} v\rangle}, \\
\label{eq:RELDEN}
\ee
where $v$ is the relative velocity and $\sigma_{ann}$ includes effects from neutralino co-annihilation with particles of similar mass.  The non-baryonic Dark Matter relic density relative to the critical density, $\Omega_{DM}$, is determined by the WMAP 3-year CMB data and the spatial distribution of galaxies to be \cite{Spergel:2006hy,Yao:2006px}
\be
\Omega_{\rm DM} h^2=0.111\pm 0.006
\label{eqn:wmap}
\ee
where $h=0.74\pm 0.03$ is the Hubble constant.  However, there could be multiple origins of dark matter, such as neutralinos and axions or non-thermal production, so strictly speaking Eq. (\ref{eqn:wmap}) provides only an upper bound on neutralino dark matter.

In this paper we consider the neutralino and Higgs states of the singlet extended supersymmetric models.  The predictions for the neutralino relic density and neutralino-nucleon scattering cross-section are often correlated.  We find that the singlet and singlino admixtures of the Higgs bosons and neutralinos, respectively, can dramatically alter the predicted relic density and the direct detection scattering cross-section.  We calculate the xMSSM predictions, along with those of the MSSM\footnote{Note that the version of the MSSM we adopt is more general than mSUGRA since we set parameters at the TeV scale.}, with the parameterization of the Higgs and neutralino sectors given in Ref. \cite{Barger:2006dh}.  Many studies on the relic density and direct detection of the lightest neutralino have been performed in the CMSSM which imposes scalar mass, gaugino mass and soft trilinear mass unification at the GUT scale, thereby reducing the number of parameters in the model \cite{Barger:1992ac,Ellis:2005mb,Baer:2003jb,Trotta:2006ew}.  Additional work on the connection between the Higgs sector and astrophysical constraints in the CMSSM have also been done in the MSSM \cite{Ellis:2005tu,Carena:2006nv}.  More general recent analyses of the MSSM can be found in Refs. \cite{Arnowitt:2007pi,Arnowitt:2006jq}.

In Section \ref{sect:ann}, we analyze the neutralino annihilation and coannihilations and how the current value of $\Omega_{DM}h^2$ constrains the Higgs and neutralino spectra.  In Section \ref{sect:scatt} we consider neutralino-nucleon scattering and the prospects for direct detection of neutralino dark matter in the xMSSM models.  Finally, in Section \ref{sect:concl}, we summarize our conclusions and provide an outlook on discovery for the models we consider. We reserve the Appendix for discussions of the neutralino mass matrix of the extended models and couplings which are altered or completely new relative to the MSSM.  Further, we discuss the constraints from the $(g-2)_\mu$ measurement and perturbativity of the couplings.

%%%%%%%%%%%%%%%%%%%%%%%%%%%%
\section{Neutralino annihilation and coannihilation}
\label{sect:ann}
%%%%%%%%%%%%%%%%%%%%%%%%%%%%

The light neutralino states in singlet extended models often have an appreciable singlino component or are dominantly singlino.  The singlino admixture has a strong influence on the neutralino annihilation cross-section and consequently on the neutralino relic density \cite{Barger:2005hb,Barger:2004bz}.  Coannihilations with supersymmetric particles of similar mass may also significantly affect the relic density.  

We calculate the relic abundance of the lightest neutralino in each model using appropriate modifications of the DarkSUSY (DS) code \cite{ds}\footnote{Other supersymmetric dark matter codes exist, including MicrOMEGAS \cite{Belanger:2006is} and ISATOOLS \cite{isajet}.}.  We change the Higgs and neutralino couplings in DS to account for the additional interactions of the singlets and singlinos (and $Z'$ inos in the UMSSM).  We list the couplings in Appendix \ref{apx:coup}.  The number of Higgs and neutralino states in DS are increased to account for the additional states in the extended models relative to the MSSM.  The Higgs boson width is recalculated for the different spectra and changes in couplings\footnote{We also include the Higgs decay to off shell gauge bosons, which can be important for intermediate Higgs masses ($120\text{ GeV} \lesssim M_{H_i} \lesssim 2 M_Z$).}.  Our calculation of the thermally averaged annihilation cross-section at tree level in our modified version of DS includes all possible initial, final, and exchanged particles of the xMSSM\footnote{Note that we do not include exotic states that can be present in some extended models.  For a list of the initial, final and exchanged states considered by DarkSUSY in the MSSM, see Table 3 in Ref. \cite{ds}.}.  For general discussion of the relic density calculation, see Ref. \cite{Jungman:1995df}.

Due to coannihilations with the extra neutralino states in xMSSM models, the relic density can be different from the MSSM result.  While Boltzmann suppression makes contributions from particles with masses $\gtrsim 1.4~m_{\N_1}$ small, we include coannihilations from particles with masses up to $2.5~m_{\N_1}$ (Ref. \cite{Edsjo:1997bg} included coannihilations up to $2.1~m_{\N_1}$).  

We fix the slepton and squark masses at a high value (5 TeV) to focus our attention on the effects the singlet has on the relic density.  Therefore, these scalar sparticles do not coannihilate with the neutralino, making their effect on the relic density negligible\footnote{Of course, additional coannihilation solutions would be possible in all models for lighter squarks and sleptons.}.  The n/sMSSM supersymmetric spectrum is typically heavy compared to the lightest neutralino making the contribution to the relic density from coannihilations negligible.  In the MSSM, NMSSM, and UMSSM, if the relic density without the effect of coannihilations is at or above the observed value of the relic density, the coannihilation contribution has little to no effect on the relic density \cite{Edsjo:1997bg}.  In this region, the neutralino is dominantly Bino, making the coupling, and therefore, the coannihilation rate with the Wino-dominated $\N_2$ and $\C_2$ small~\footnote{Again, assuming the sleptons are heavy, thereby suppressing the coannihilation process $\N_1 \N_i \to f\bar f$.}.

\begin{figure}[tb]
\centering
\includegraphics[width=0.35\textwidth,angle=-90]{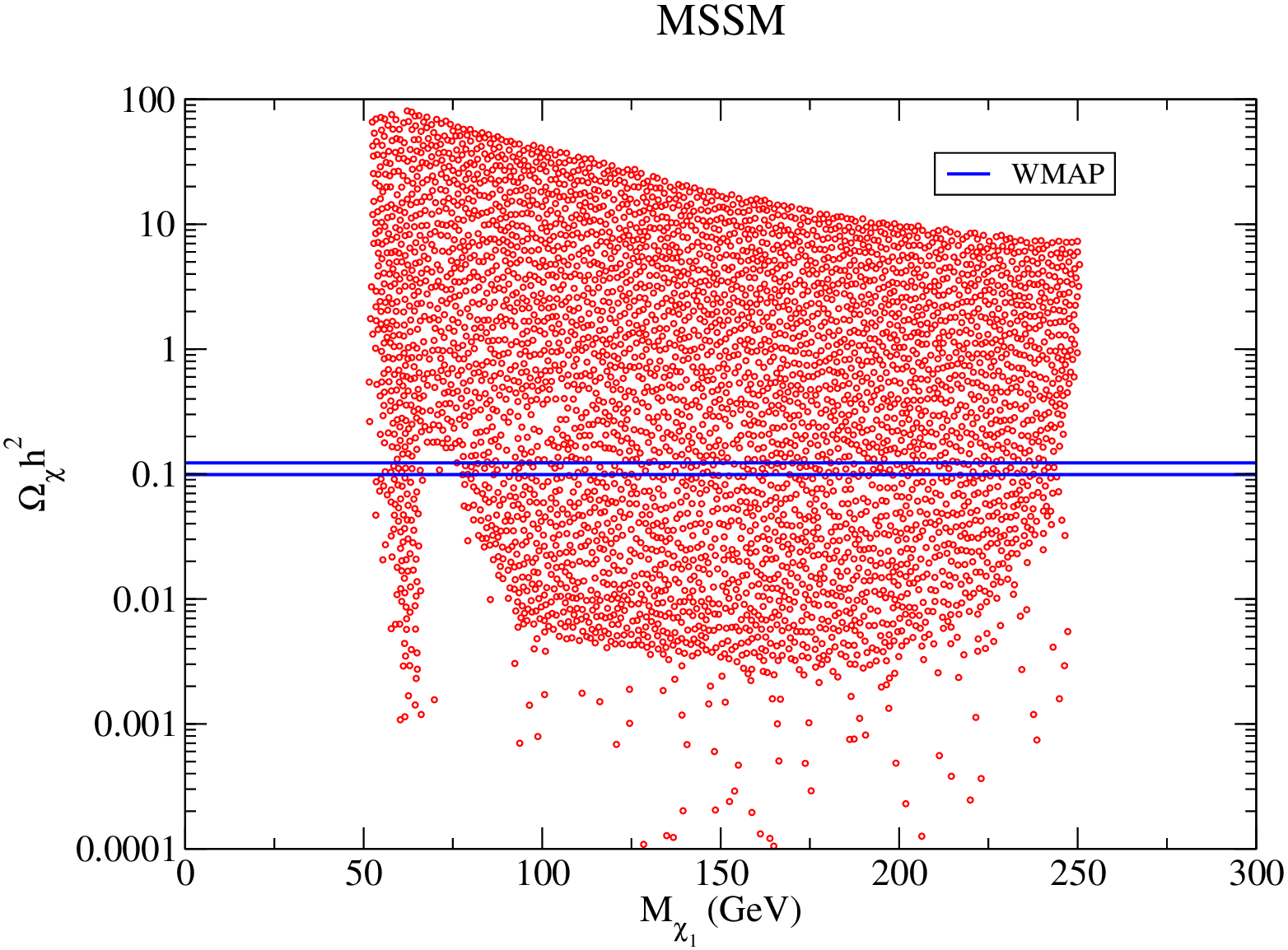}
\includegraphics[width=0.35\textwidth,angle=-90]{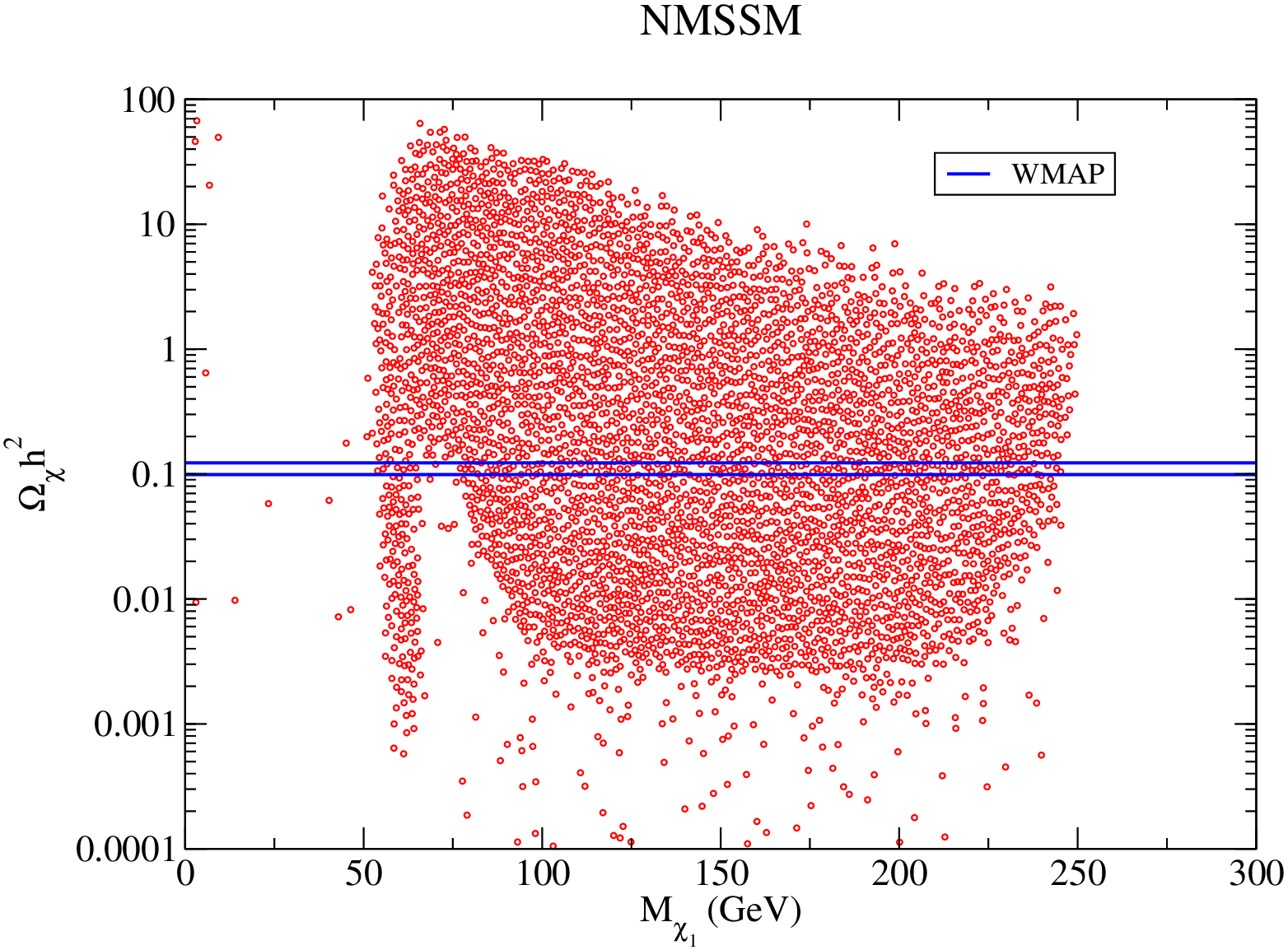}
\includegraphics[width=0.35\textwidth,angle=-90]{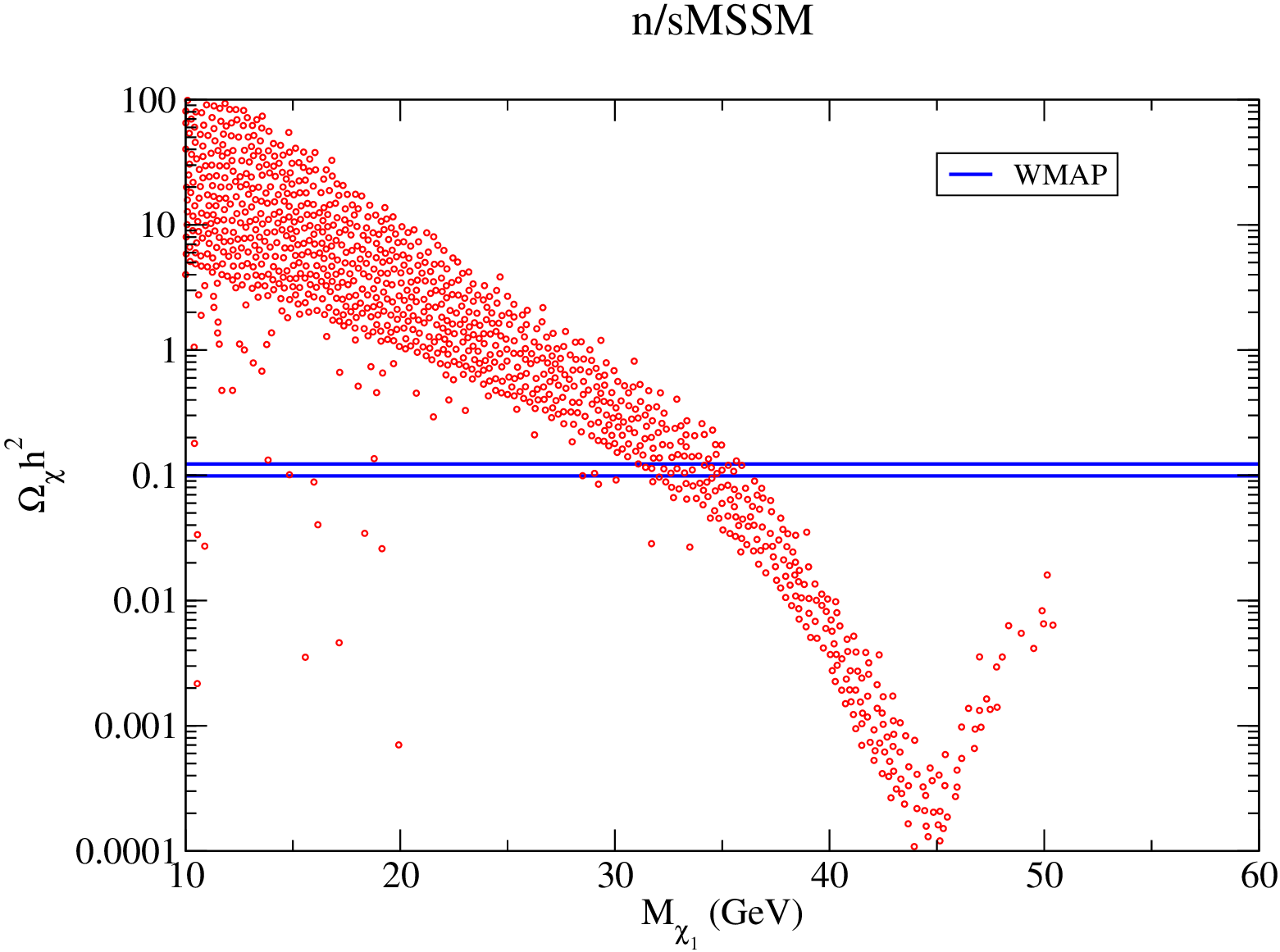}
\includegraphics[width=0.35\textwidth,angle=-90]{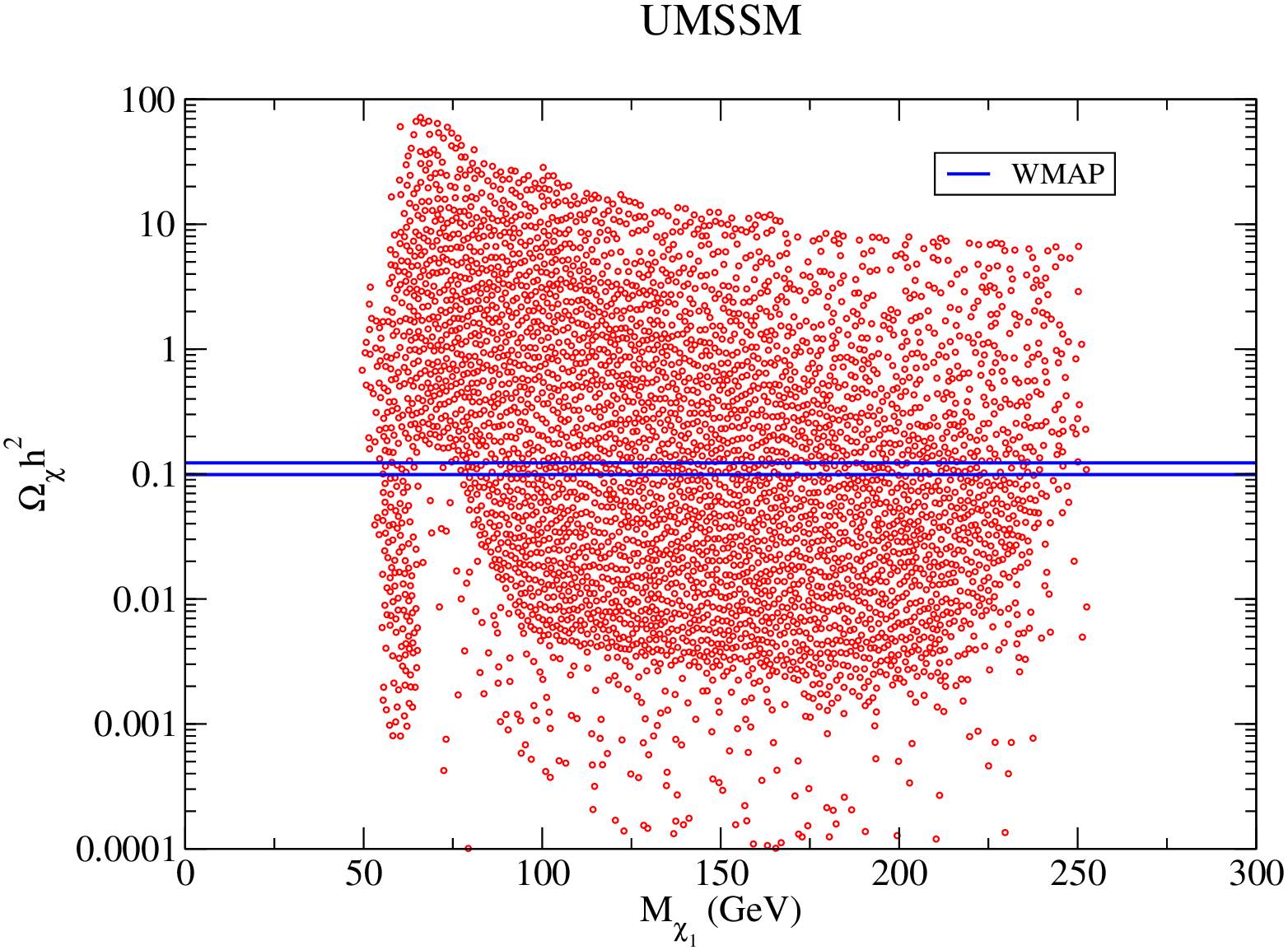}
\caption{Neutralino relic density versus the lightest neutralino mass.  The relic density is constrained to be in the region $0.123>\Omega_{DM} h^2>0.099$ provided that the model is solely responsible for the observed dark matter.  The efficient annihilations through the Higgs boson pole in the MSSM, NMSSM and UMSSM are evident at $m_{\N_1} \sim M_{H_1}/2 \sim 60$ GeV and of the $Z$ boson pole at $m_{\N_1}\sim M_Z / 2$ in the n/sMSSM.}
\label{fig:RELDEN}
\end{figure}

The calculated relic density of the neutralino in the various models is shown in Fig. \ref{fig:RELDEN} versus the mass of $\N_1$.  The region $0.123>\Omega_{DM} h^2>0.099$ is the $2\sigma$ observed range \cite{Spergel:2006hy,Yao:2006px}.   Values $\Omega_{\N_1} h^2<0.099$ are allowed in the event that other particles contribute to the relic abundance of dark matter, or there is an enhanced non-thermal production mechanism.

The profiles of the relic density versus the lightest neutralino mass are similar for the MSSM, NMSSM, and UMSSM due to their similar low energy neutralino spectra.  In the NMSSM and UMSSM, the singlino is typically substantially heavier than the lightest neutralino, $\N_1$, preventing a large singlino mixture in $\N_1$.  However, the NMSSM allows a lower neutralino mass than in the UMSSM and MSSM, as the $\kappa \to 0$ limit resembles the n/sMSSM which has a very light $\N_1$.

Overall, there is a broad filled region in Fig. \ref{fig:RELDEN} for the MSSM, NMSSM and UMSSM where the relic density varies depending on the composition and mass of the lightest neutralino.  For gaugino dark matter, the annihilation rate is not strong enough to yield the observed dark matter.  As the higgsino content increases, the relic density decreases as the annihilation rate becomes larger.  The suppression of the relic density due to efficient annihilation through the $H_1$ pole near $m_{\N_1}\sim 60$ GeV is evident.  The $H_2$ pole may also have a similar effect, as indicated by the sporadic points falling below the filled regions.  In this low region, namely $\Omega_{\N_1} h^2 \lesssim \text{ few }\times 10^{-3}$, the NMSSM and UMSSM have more points than the MSSM since there are more Higgs resonances that contribute.

\begin{figure}[htbp]
\begin{center}
\includegraphics[width=0.35\textwidth,angle=-90]{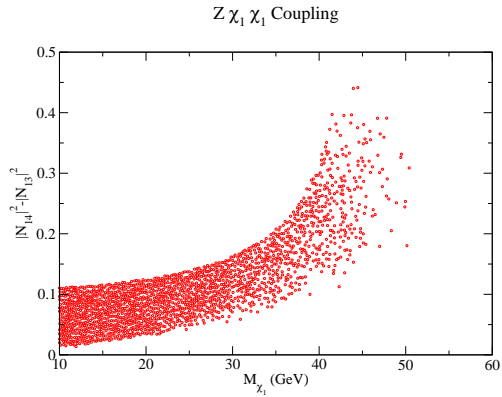}
\caption{The $Z\N_1\N_1$ coupling ($\propto |N_{14}|^2-|N_{13}|^2$) decreases in the n/sMSSM as the neutralino mass decreases, assisting in the enhancement of the relic density in the low mass region.}
\label{fig:zcc}
\end{center}
\end{figure}

For the n/sMSSM, the lightest neutralino mass is usually below 50 GeV \cite{Barger:2004bz,Menon:2004wv}.  Annihilation through the $Z$ boson pole dominates the rate near $m_{\chi^0_1}\approx 45$ GeV, decreasing the relic abundance below the observed value.  Additionally, in the n/sMSSM the relic density is strongly dependent on the mass of the lightest neutralino.  As the neutralino mass decreases, the annihilation rate becomes suppressed by the $Z$ propagator, increasing the relic density.  Furthermore, in the n/sMSSM, there is a $Z \N_1\N_1$ coupling suppression since the neutralino increasingly becomes singlino at lower $m_{\N_1}$ as shown in Fig. \ref{fig:zcc}.  Therefore, an approximate lower bound on the lightest neutralino mass in the n/sMSSM can be placed at $m_{\N_1}\gtrsim 30$ GeV.

Some points in the n/sMSSM exist below the 30 GeV $m_{\N_1}$ bound where the relic density falls at or below the observed value due to the contribution of additional channels.  For example, a light $A_1$ resonance enhances the annihilation cross-sections.  Similarly, the $A$-funnel region in mSUGRA parameter space \cite{Lahanas:2001yr,Ellis:2001ms} can account for the observed relic density.  However, in the present case, the lightest CP-odd Higgs is singlet.  Excluding annihilation through the $A$-pole, the neutralino mass range compatible with the lower and upper bounds of $\Omega_{DM}$ are $30~\text{GeV}~\lesssim m_{\chi^0_1} \lesssim 37$ GeV~\footnote{Since we only consider sfermion masses much heavier than the neutralino, coannihilation with these particles do not contribute to the effective annihilation rate.  In general slepton and squark coannihilation can be important as we parameterize the MSSM more generally than mSUGRA.}.  The limit might also be slightly weakened in the full secluded model \cite{smssm,Han:2004yd} which may include a singlino $\N_0$ even lighter than the $\N_1$.  Even a tiny $\N_0-\N_1$ mixing, irrelevant for collider physics, could allow $\N_1$ to decay, e.g., to $\N_0 f \bar f$, reducing $\Omega_{\N}h^2$ by $m_{\N_0}/m_{\N_1}$ \cite{Feng:2004gg}.

\begin{figure}[tb]
\centering
\includegraphics[width=0.35\textwidth,angle=-90]{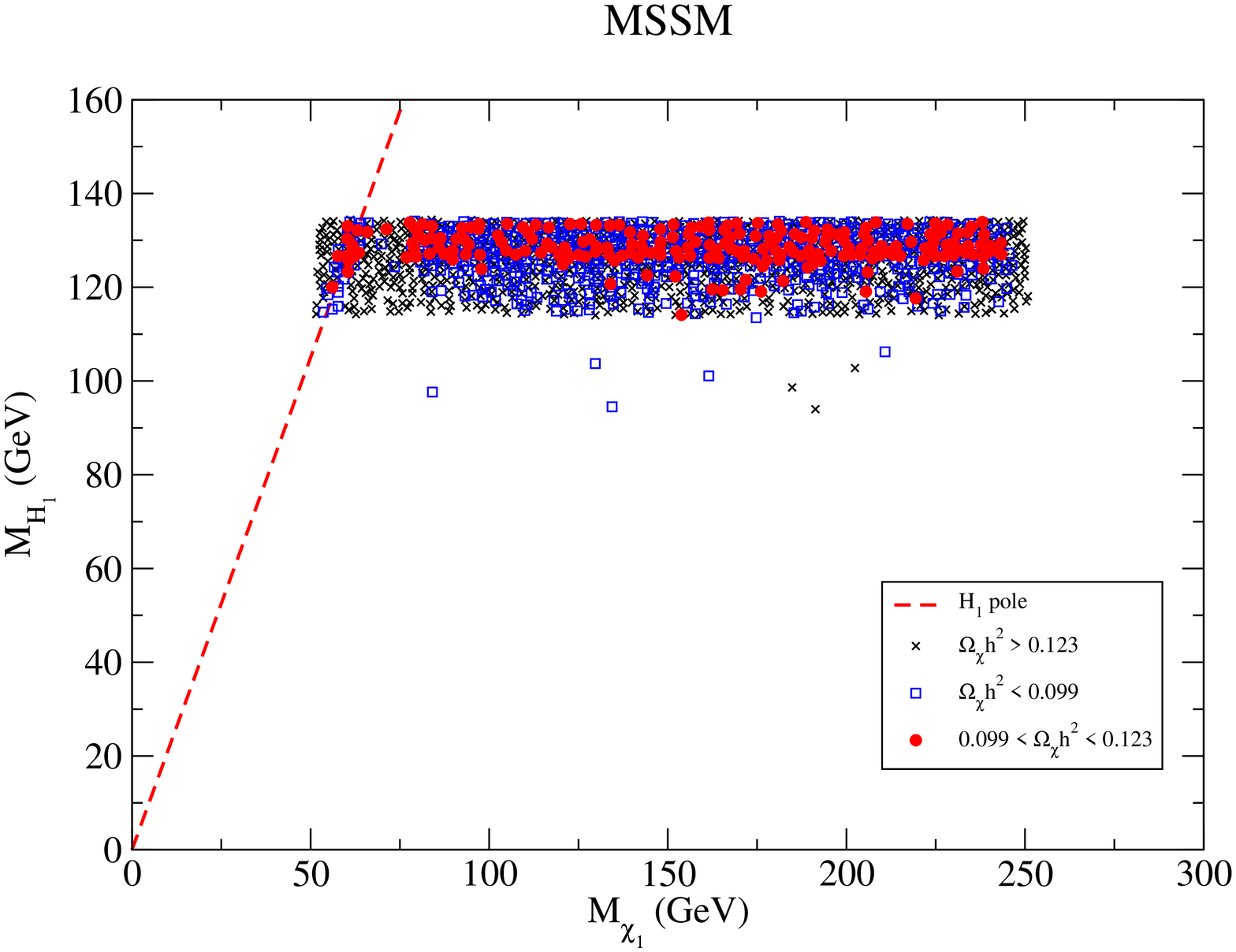}
\includegraphics[width=0.35\textwidth,angle=-90]{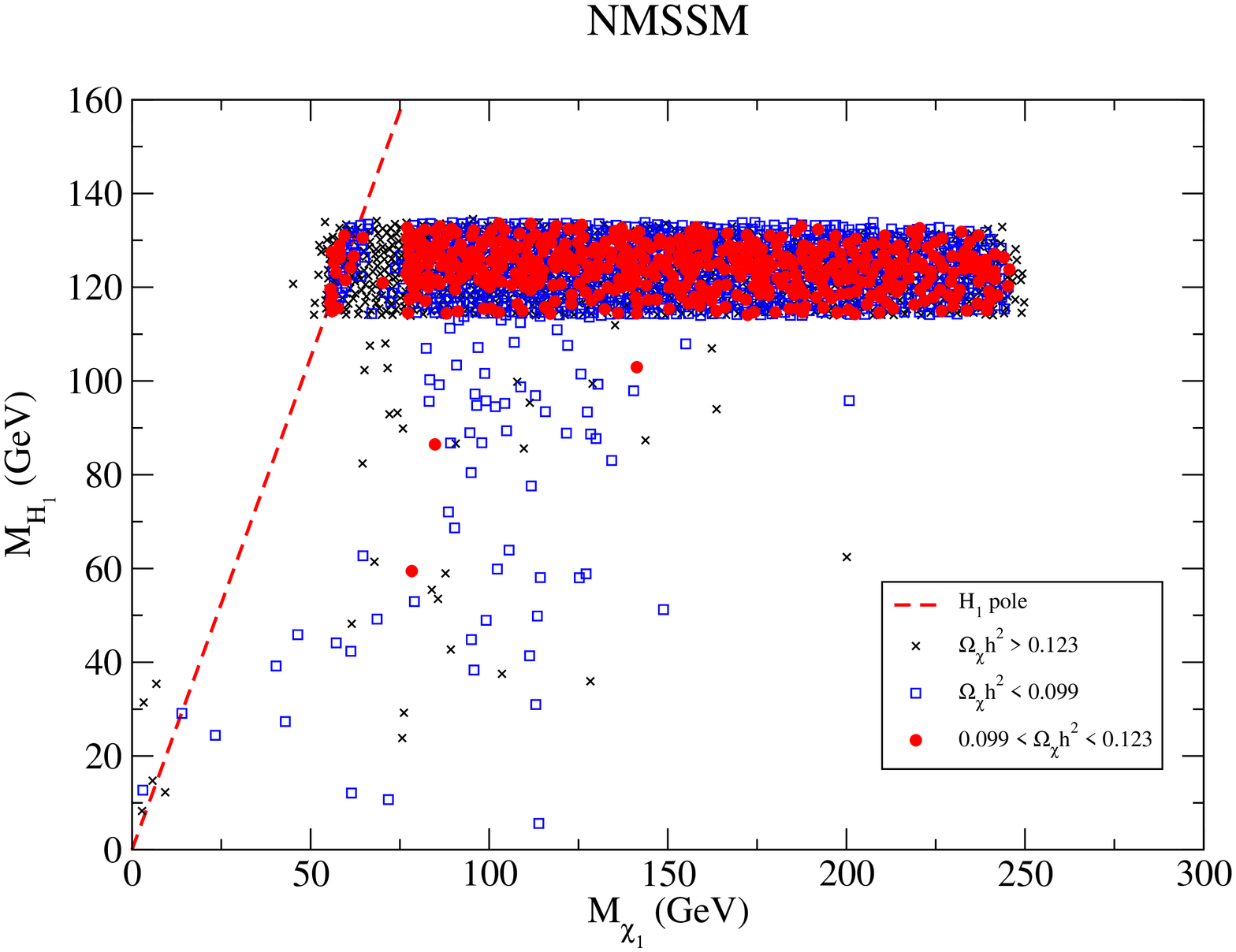}
\includegraphics[width=0.35\textwidth,angle=-90]{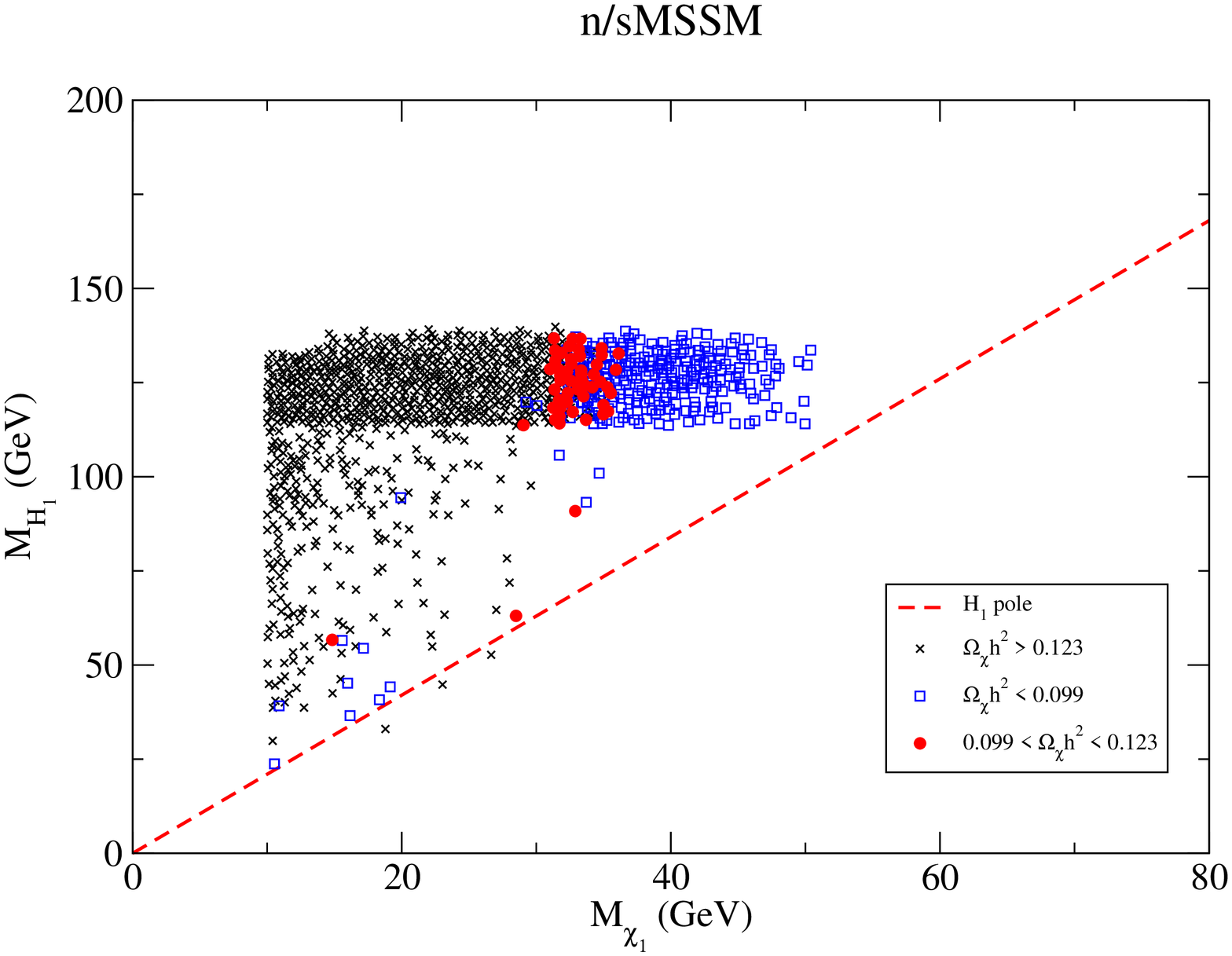}
\includegraphics[width=0.35\textwidth,angle=-90]{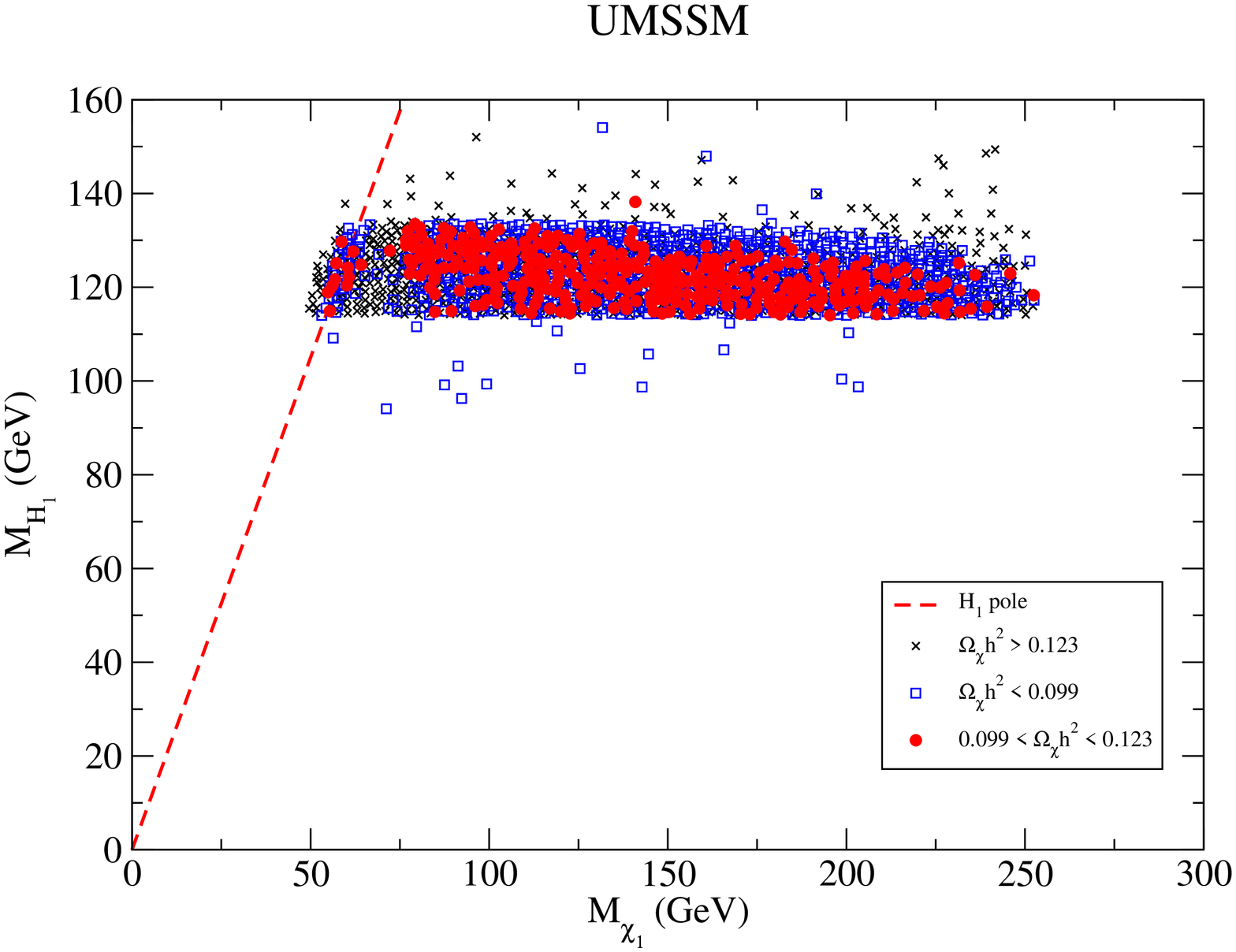}
\caption{Lightest Higgs mass versus lightest neutralino mass for the MSSM, NMSSM, n/sMSSM, and UMSSM.  An over-density of relic neutralinos is denoted by a  black x, an under-density is denoted by a blue box and the WMAP observed density is denoted by a red circle.}
\label{fig:mchimh}
\end{figure}

To further document the effect of the $H_1$ pole on the relic density, we plot in Fig. \ref{fig:mchimh} the masses of the lightest Higgs boson and neutralino for various ranges of the relic density.  In models with singlet mixing, the lightest Higgs mass can be lower than in the MSSM, filling the major band of $114\text{ GeV} < M_{H_1} < 135\text{ GeV}$.

In Fig. \ref{fig:mchimh}, the MSSM, NMSSM, and UMSSM show bands of allowed points associated with the $H_1$ pole near the line $m_{\N_1} \simeq 2.1 m_{H_1}$.  The overdense region where $m_{\N_1} \sim 75$ GeV is due to inefficient annihilation of gaugino-like neutralinos.  In the NMSSM, there are points with low allowed $H_1$ masses near this band; most points there are due to annihilation through a SM-like $H_2$.  

The n/sMSSM shows a weak dependence on the Higgs mass since the most dominant annihilation is through the $Z$ boson.  The vertical bands of accepted points are due to the dependence of the relic density on the lightest neutralino mass, as discussed above.

\begin{figure}[t]
\centering
\includegraphics[width=0.35\textwidth,angle=-90]{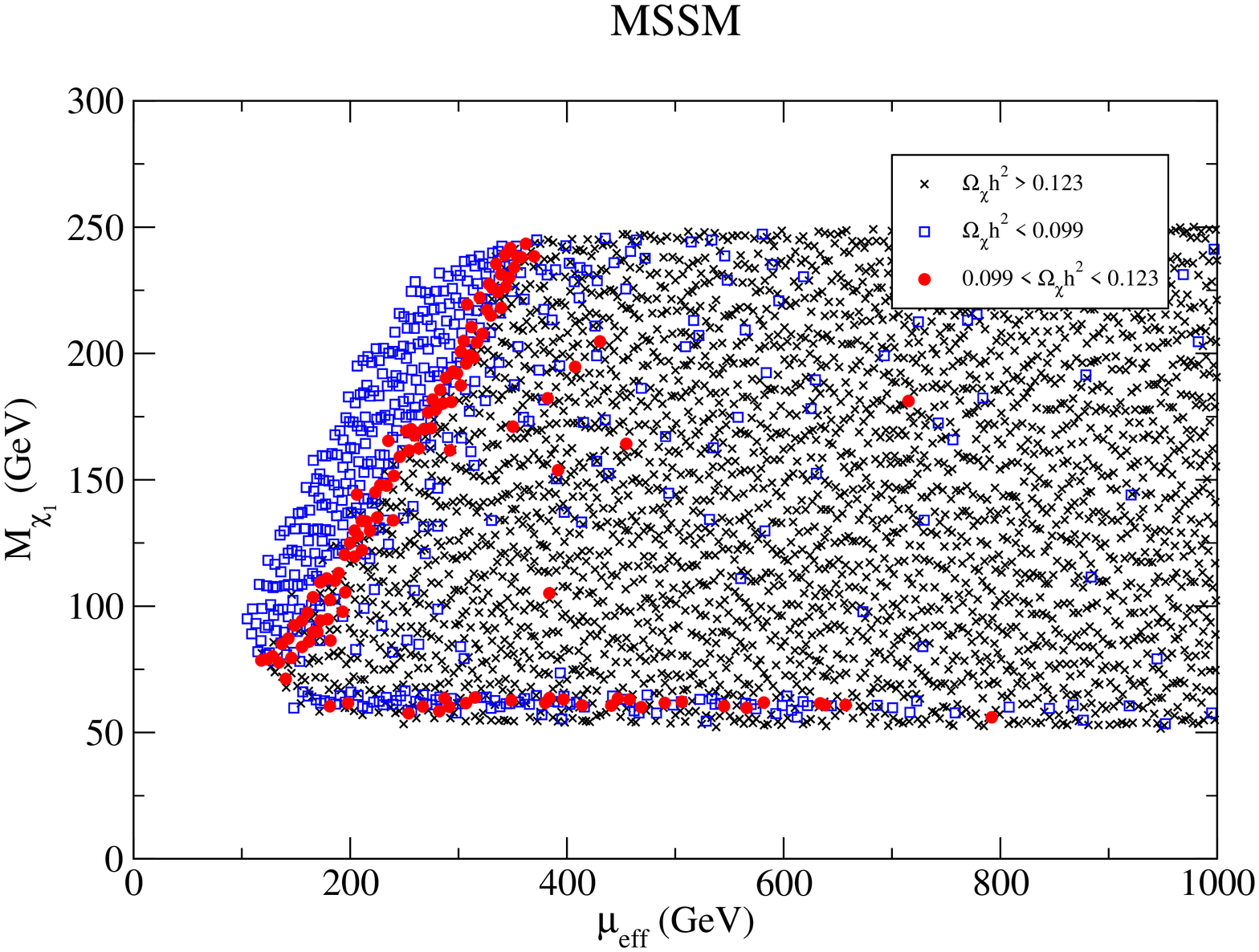}
\includegraphics[width=0.35\textwidth,angle=-90]{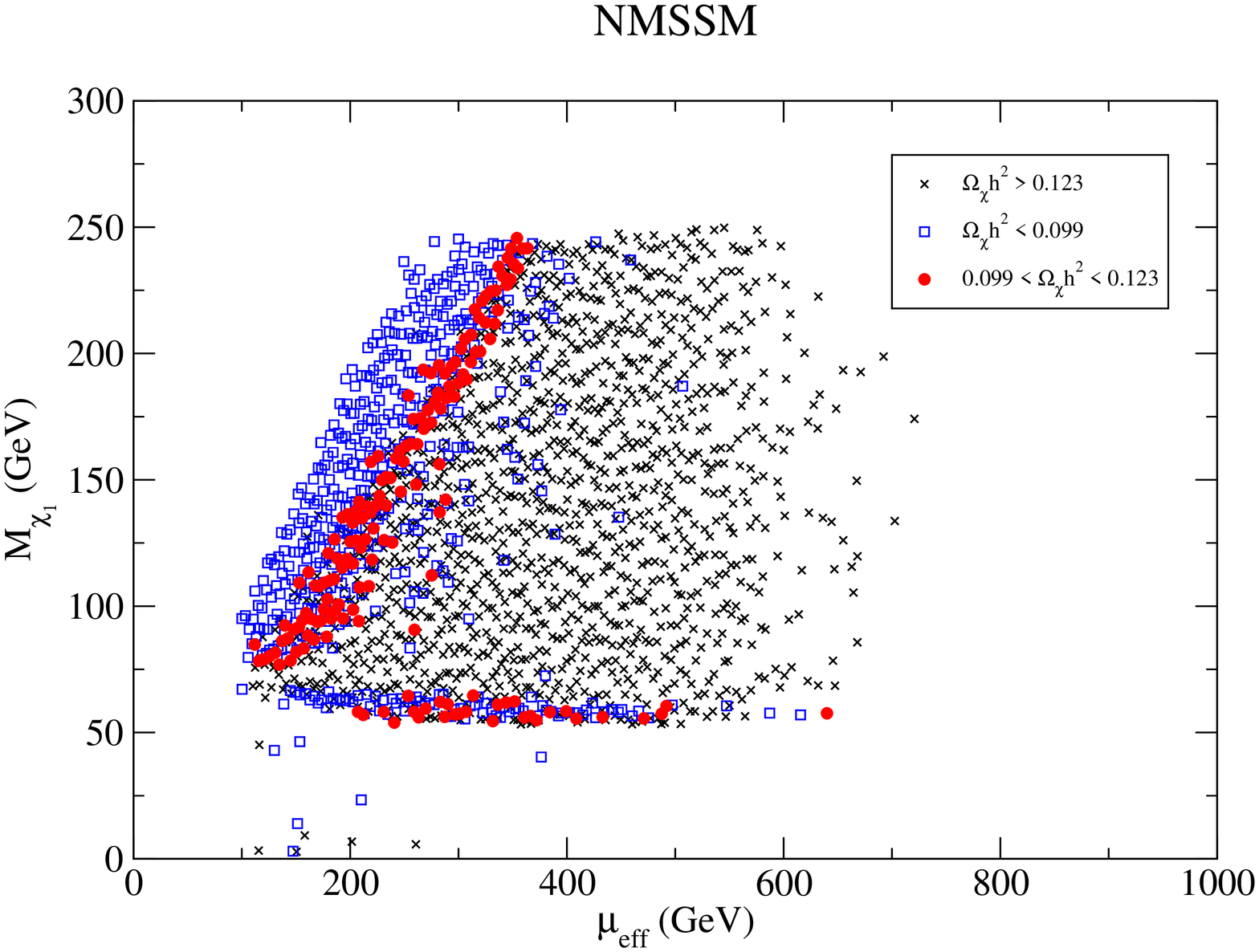}
\includegraphics[width=0.35\textwidth,angle=-90]{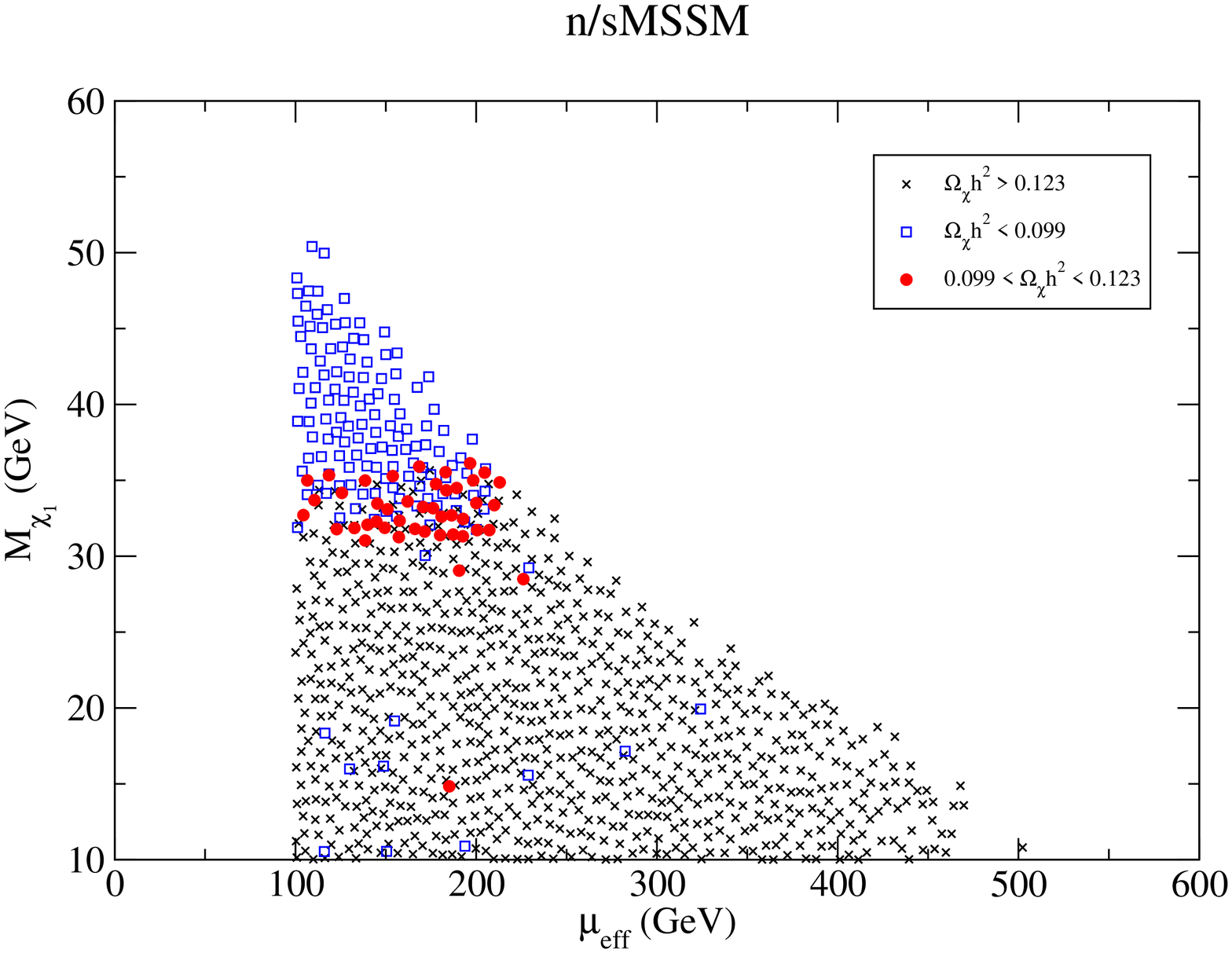}
\includegraphics[width=0.35\textwidth,angle=-90]{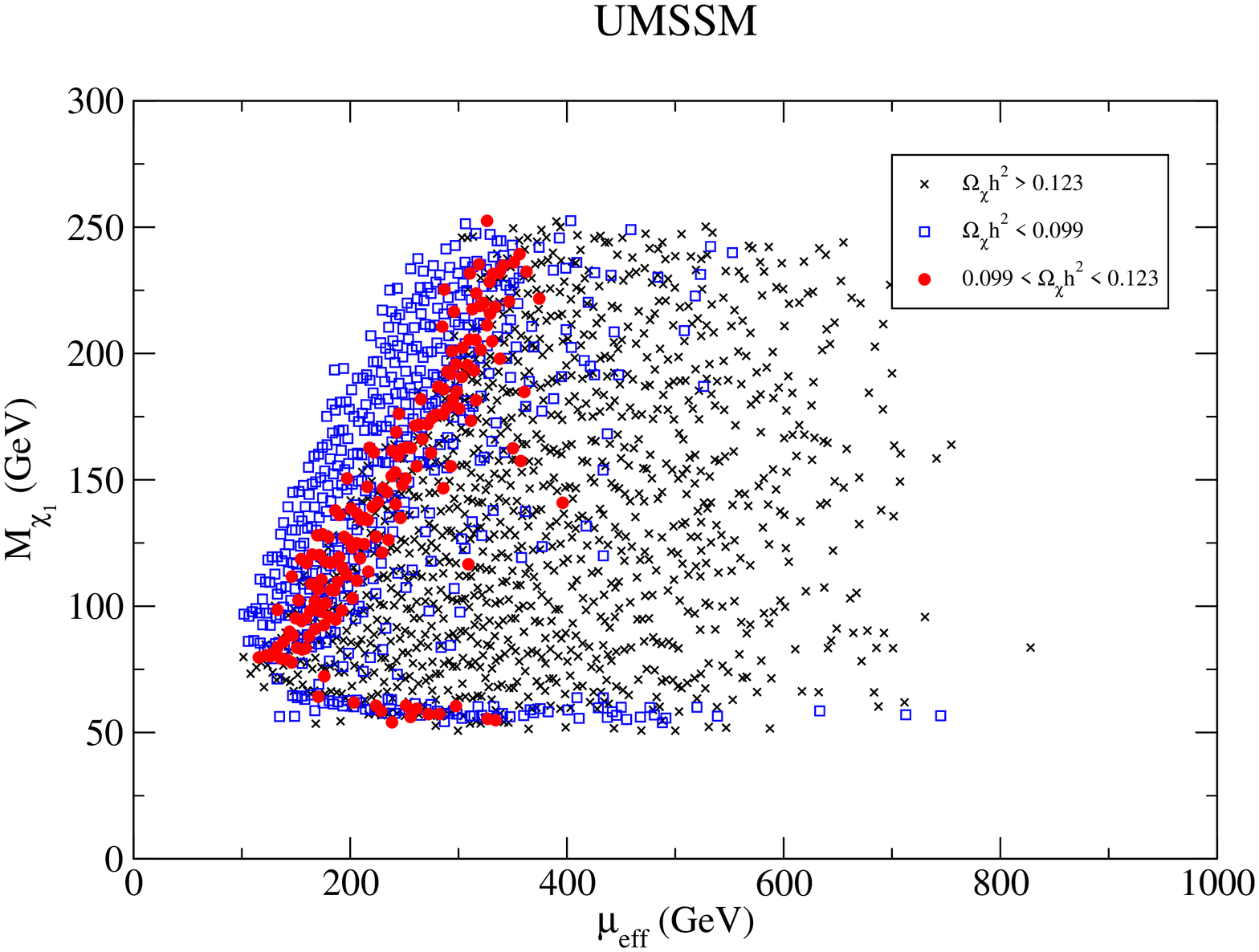}
\caption{Regions of varying relic density in $\mueff$ vs. $m_{\N_1}$.  The preferred horizontal strip in the MSSM, NMSSM and UMSSM at $m_{\N_1}\sim 60$ GeV is due to the Higgs pole.  The steeply sloped region corresponds to the focus point.}
\label{fig:mchimu}
\end{figure}

Fig. \ref{fig:mchimu} shows the relic density in the plane of $\mueff$ and the lightest neutralino mass.  Two bands of allowed relic density values are evident in the MSSM, NMSSM and UMSSM.  The horizontal band at $M_H\sim 60$ GeV is associated with the lightest Higgs pole, while the other band is the focus point region \cite{Feng:2000gh,Baer:2005ky}.  This region is characterized by a Higgsino-bino mixed neutralino  which makes neutralino annihilation efficient enough to reproduce the observed relic density \cite{Baer:2005ky}.  Note that even though the focus point region requires heavy sleptons and mixed higgsino dark matter, the neutralino may still coannihilate with neutralinos and charginos of similar mass.  Increasing $\mueff$ above this region makes the neutralino more gaugino-like, which decreases the annihilation rate.  However, decreasing $\mueff$ will make relic neutralinos annihilate more efficiently, resulting in a relic density below the observed $\Omega_{DM}h^2$ band.  The n/sMSSM has a neutralino that is dominantly singlino, so the relic density is insensitive to $\mueff$.

\begin{figure}[tb]
\centering
\includegraphics[width=0.35\textwidth,angle=-90]{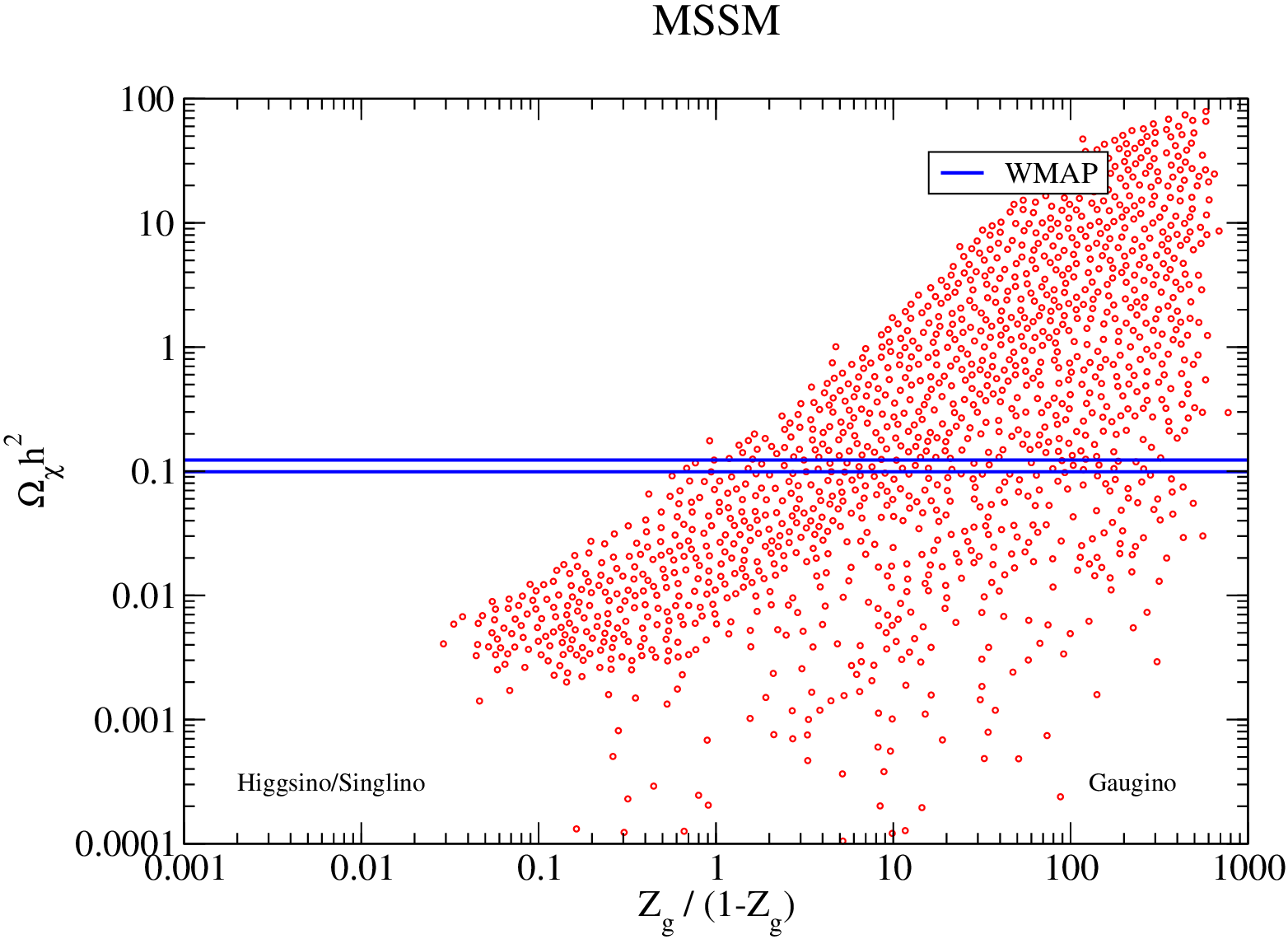}
\includegraphics[width=0.35\textwidth,angle=-90]{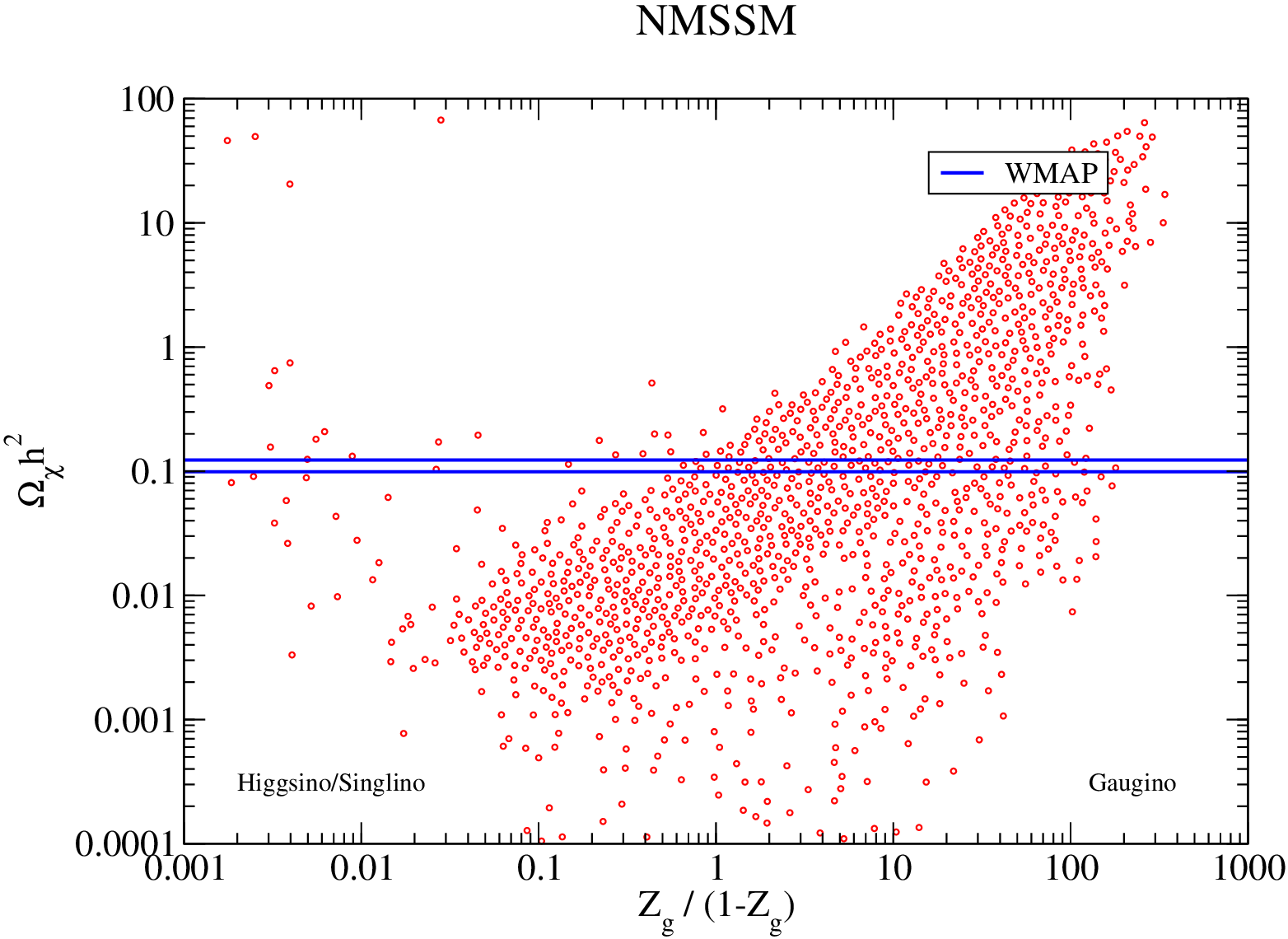}
\includegraphics[width=0.35\textwidth,angle=-90]{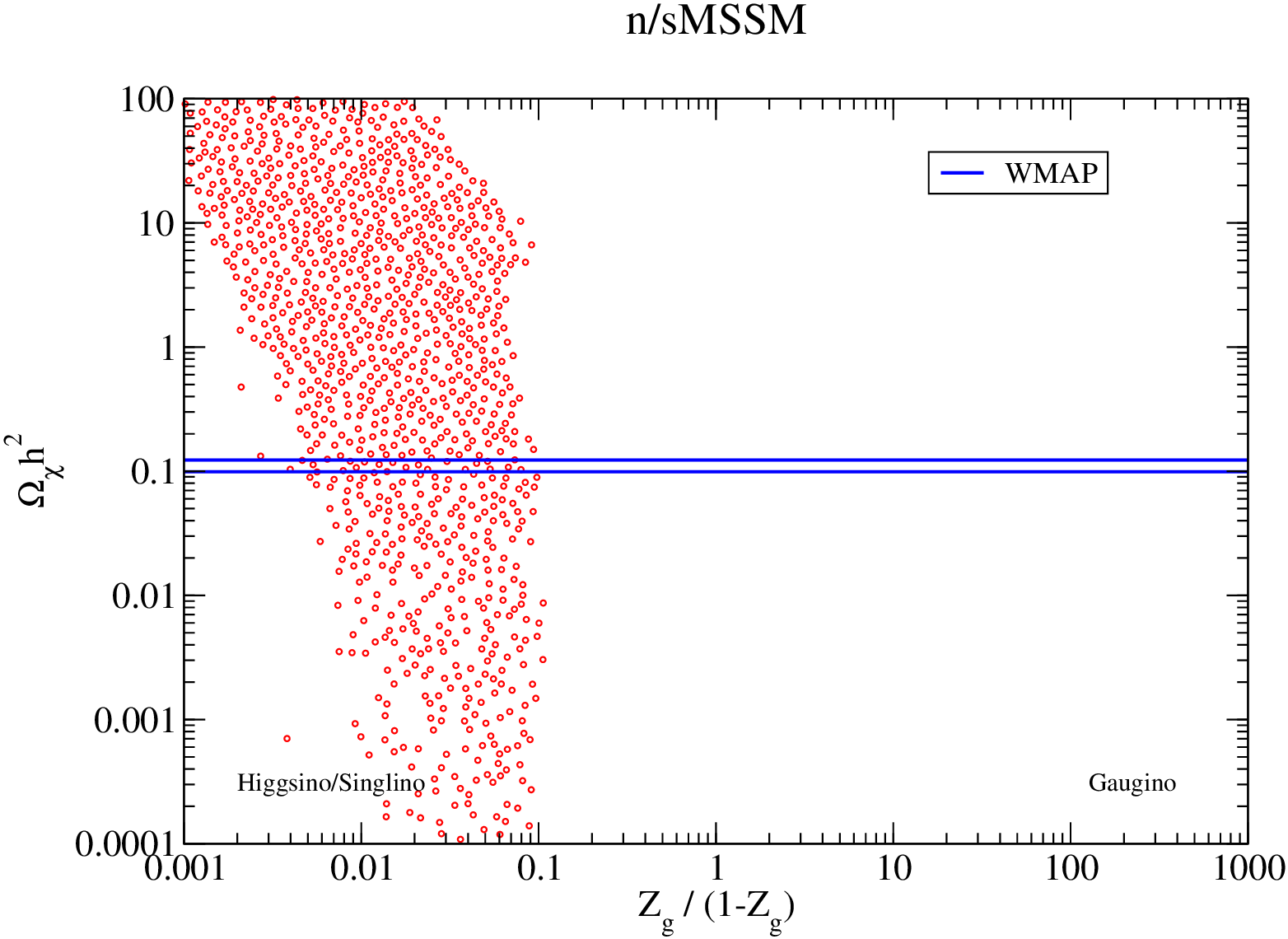}
\includegraphics[width=0.35\textwidth,angle=-90]{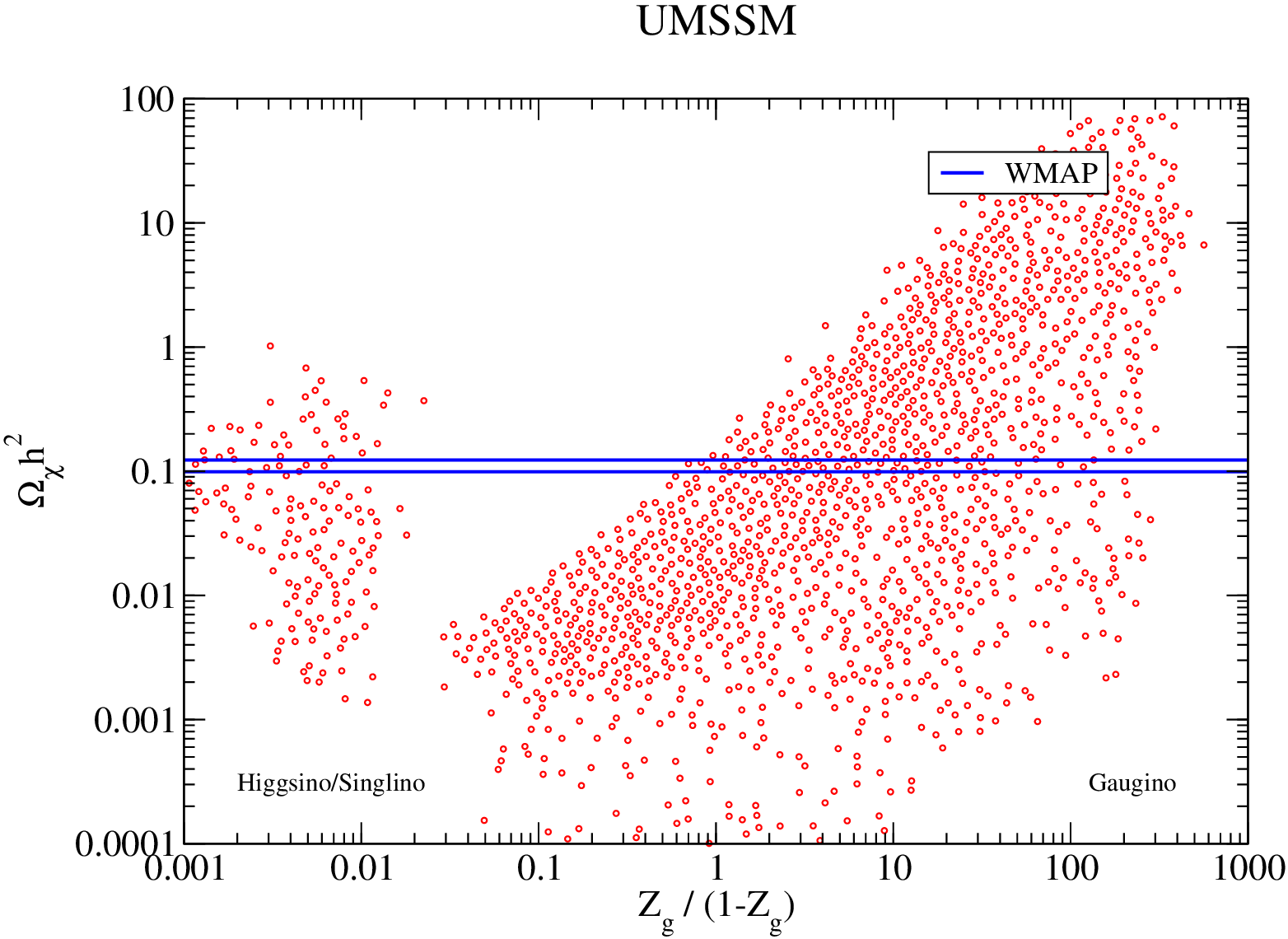}
\caption{Neutralino relic density dependence on the MSSM gaugino fraction $Z_g$.  Points at the left end of each panel have a lightest neutralino that is dominantly higgsino and singlino, while at the right side it is dominantly gaugino.}
\label{fig:gfrac}
\end{figure}

The gaugino composition of the lightest neutralino is shown in Fig. \ref{fig:gfrac}, where $Z_g=|N_{11}|^2+|N_{12}|^2$ is the MSSM gaugino fraction.  The right sides of the panels indicate a large gaugino fraction while the left sides indicate neutralinos with high Higgsino or singlino/$Z'$-ino composition.  In the MSSM, the broad band of WMAP allowed points between $1 \lesssim Z_g / (1-Z_g) \lesssim 10$ correspond to the focus point region while the other scattered points with higher gaugino fraction are due to the Higgs poles discussed earlier.  

\begin{figure}[tb]
\centering
\includegraphics[width=0.35\textwidth,angle=-90]{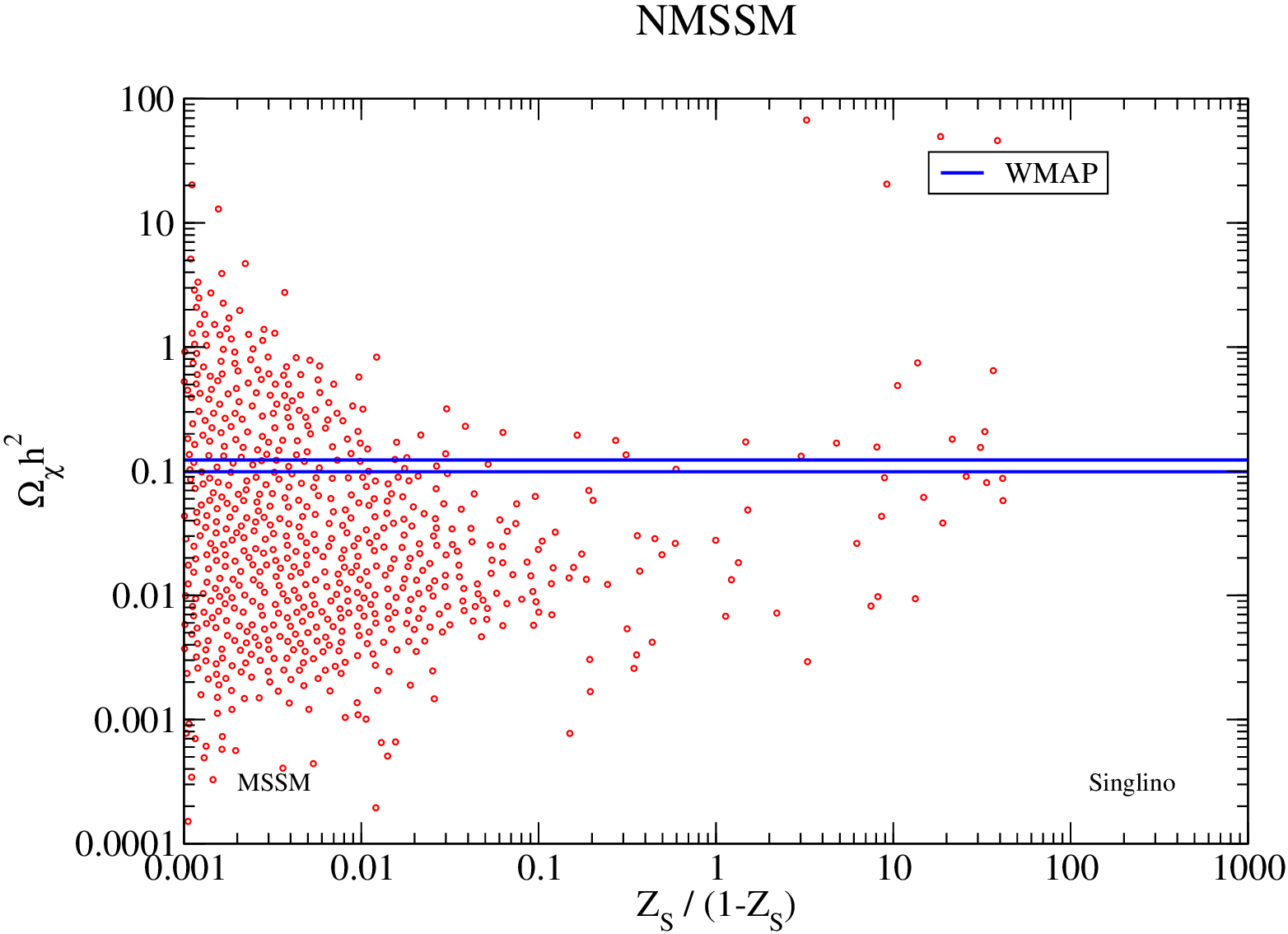}
\includegraphics[width=0.35\textwidth,angle=-90]{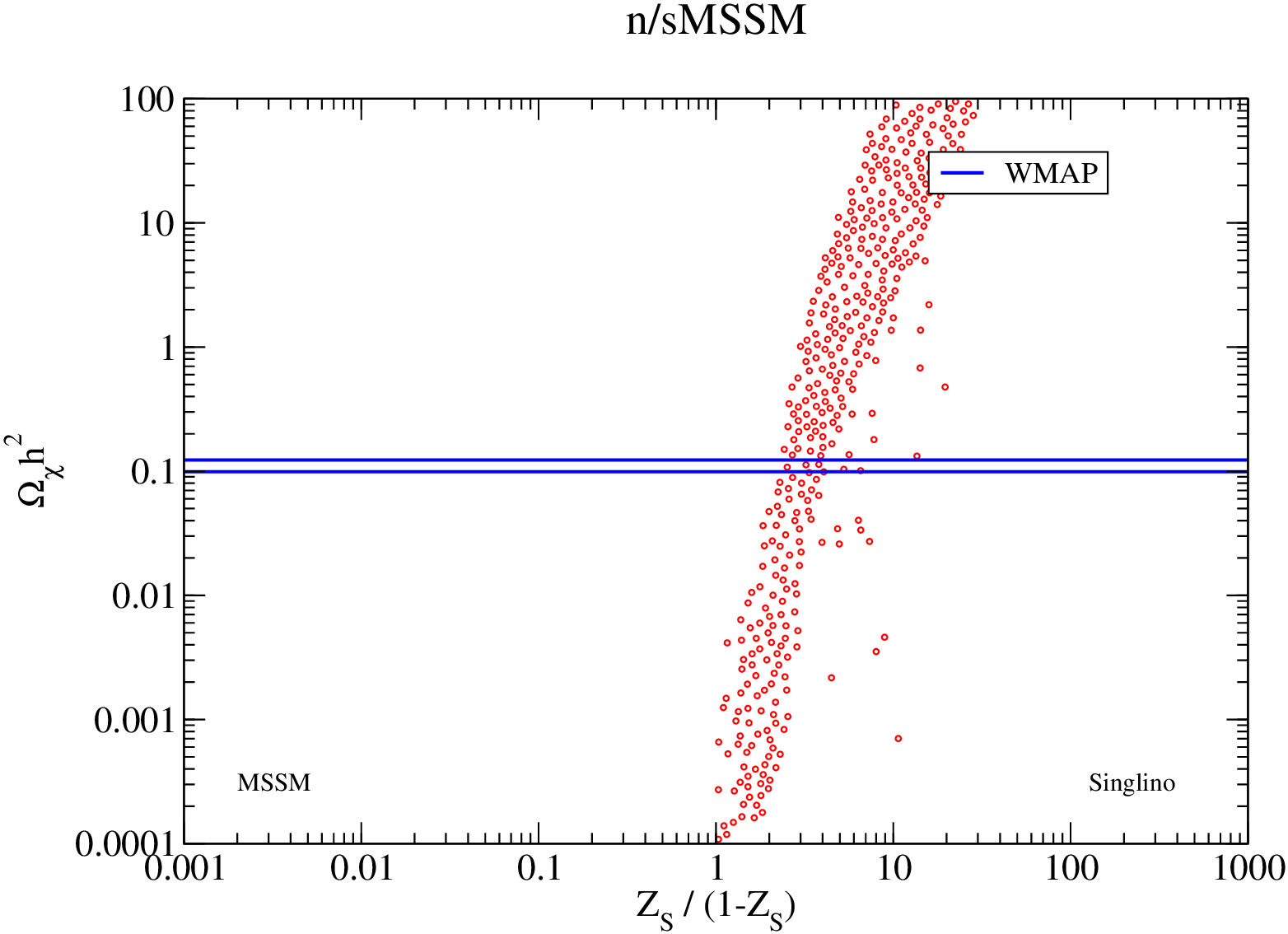}
\includegraphics[width=0.35\textwidth,angle=-90]{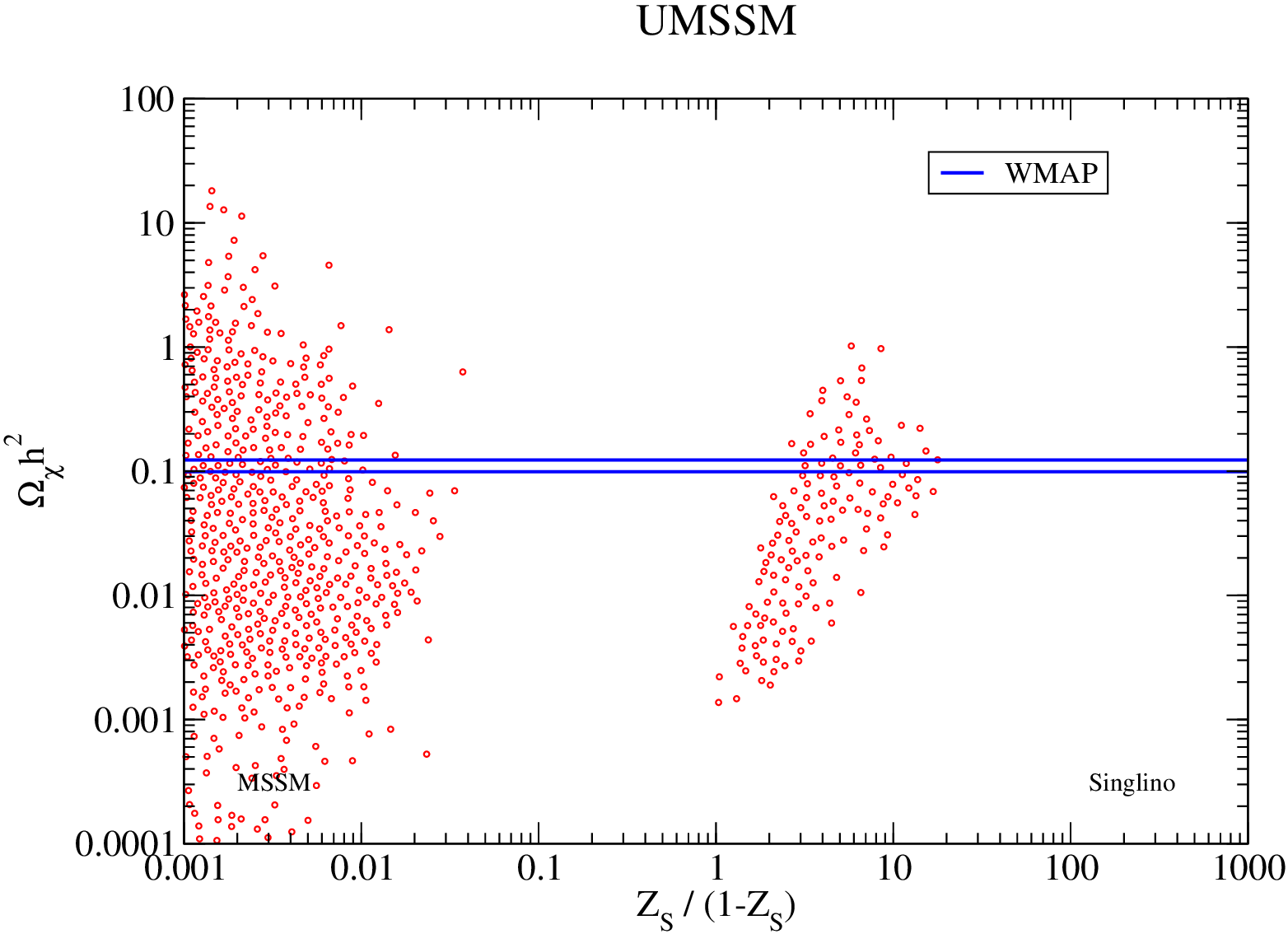}
\caption{Neutralino relic density dependence on the singlino/$Z'$-ino fraction $Z_S$.  Points at the left end of each panel have a lightest neutralino that is MSSM-like, while at right side it is dominantly singlino or $Z'$-ino.}
\label{fig:sfrac}
\end{figure}

In Fig. \ref{fig:sfrac}, we show the singlino and $Z'$-ino compositions of the neutralino.  The UMSSM and n/sMSSM both have a range of singlino/$Z'$-ino fractions that are dominant, but not close to maximal.  As the neutralino becomes less MSSM-like, the relic density increases.  

Since the coupling between the neutralino pairs and SM particles ($Z$ and Higgs bosons) needs a suitable value to give an acceptable annihilation rate, the lightest neutralino requires a nonzero MSSM fraction\footnote{Since we assume that the sleptons are heavy with masses at the TeV scale, the contributions of $t$-channel diagrams are suppressed.}.  In the n/sMSSM, the annihilation is dominated by the $Z$ boson, so the lightest neutralino requires a non-negligible Higgsino component.  

The singlino and $Z'$-ino dominated neutralino in the UMSSM can also explain for the observed relic density.  The coupling between the lightest neutralino pair and the singlet dominated $H_2$ can be as strong as the $\N_1\N_1 H_1$ coupling in the MSSM, c.f. Eq. (\ref{eqn:xxh}).  The strong coupling and resonant enhancement can yield a large enough annihilation rate to be below the $\Omega_{DM}$ observation.  However, there are parameter points which do fit the observed relic abundance.

%%%%%%%%%%%%%%%%%%%%%%%%%%%%
\section{$\N_1 p \to \N_1 p$ elastic scattering}
\label{sect:scatt}
%%%%%%%%%%%%%%%%%%%%%%%%%%%%

The singlet extended models can give significant changes in the predicted MSSM cross-sections relevant to future recoil direct detection experiments.  We use the modified version of DarkSUSY discussed in Section \ref{sect:ann} to calculate the spin-independent (scalar interaction) and spin-dependent (vector interaction) elastic scattering cross-sections of the lightest neutralino off nucleons.  

The experimental sensitivity to spin-independent (SI) scattering is much larger since spin-independent processes scatter coherently, and therefore are enhanced in scattering from large target nuclei.  However, spin-dependent (SD) measurements can be made and have been probed with the Cryogenic Dark Matter Search (CDMS) to the few pb level for elastic proton scattering \cite{Akerib:2005za}.  The Chicagoland Observatory for Particle Physics (COUPP) experiment which uses superheated $CF_3 I$ can greatly improve the limits on the SD processes, down to $10^{-2}$ pb for proton scattering for a 2kg chamber, which is close to the upper cross-section expected in the MSSM \cite{Bolte:2006pf}.

Current detection experiments such as EDELWEISS \cite{Sanglard:2005we} and CDMS have SI sensitivities on the order of $10^{-6}$pb.  In 2007 the sensitivity of the CDMS II experiment is expected to improve to nearly $10^{-8}$pb \cite{Akerib:2006ri}.  The proposed future upgrade, SuperCDMS, would reach a detection sensitivity of $10^{-9}$pb.  The WARP experiment with warm liquid Argon is projected to reach a sensitivity of $10^{-10}$pb and below; this experiment has just reported first results \cite{Benetti:2007cd}.  The WARP sensitivity of an initial run with 2.3L and a total fudicial exposure of 96.5 kg$\cdot$day are slightly better than that obtained with EDELWEISS.  For recent reviews on direct detection experiments see Refs. \cite{Ellis:2005mb,Trotta:2006ew,Baer:2003jb} and for a forthcoming comprehensive summary of the status of direct detection experiments, see Ref. \cite{DMSAG}.

The SI scattering cross-section of a neutralino off a nucleus is given by \cite{ds}
\be
\sigma^{SI}_{\chi_{i}}=\frac{\mu^{2}_{\chi_{i}}}{\pi}|ZG^{p}_{s}+(A-Z)G^{n}_{s}|^{2}
\ee
where $\mu_{\chi^{i}}={m_{\N_i}m_N\over m_{\N_1} + m_N}$ is the reduced nucleon-neutralino mass.  The parameters $G_s^p$ and $G_s^n$ are hadronic matrix elements.  Since the cross-sections for scattering off protons and neutrons are very similar, we focus on scattering from protons, for which the cross-section is
\be
\sigma^{SI}_{\chi p}=\frac{\mu^{2}_{\chi p}}{\pi}|G^{p}_{s}|^{2}
\ee

\begin{figure}[tb]
\begin{center}
\includegraphics[width=0.19\textwidth]{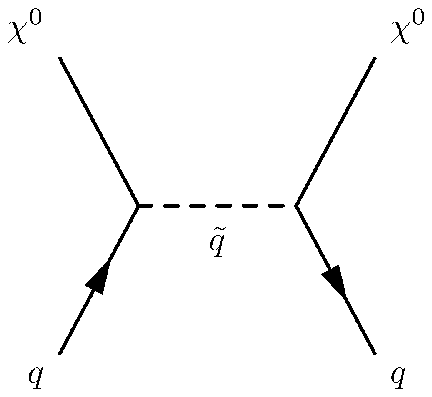}
\includegraphics[width=0.19\textwidth]{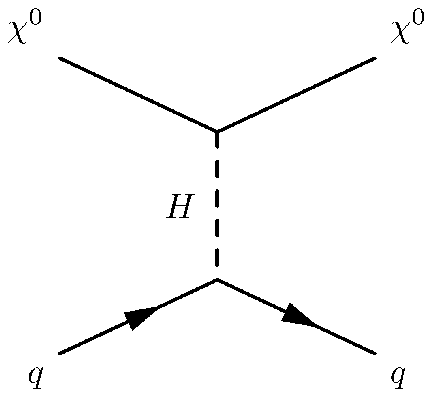}
\includegraphics[width=0.19\textwidth]{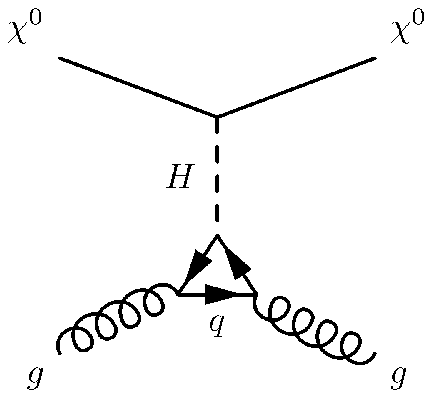}
\includegraphics[width=0.19\textwidth]{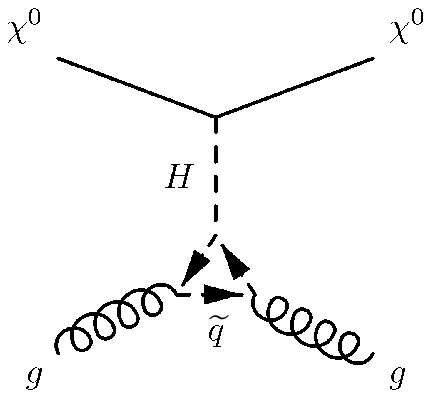}
\includegraphics[width=0.19\textwidth]{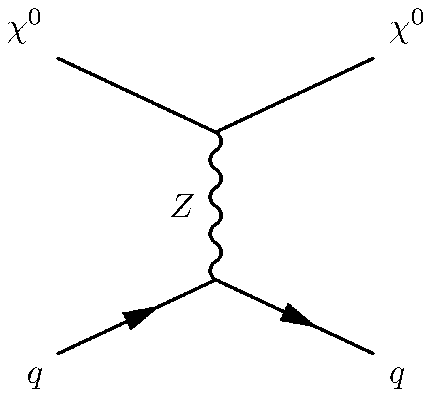}\\
(a)\hspace{.17\textwidth}(b)\hspace{.17\textwidth}(c)\hspace{.17\textwidth}(d)\hspace{.17\textwidth}(e)
\caption{Diagrams contributing to neutralino-hadron scattering.  Scalar quark (a), Higgs boson (b) and $Z$ boson (e) exchange contribute to the SI scattering cross-section.  A significant contribution also comes from the Higgs scattering off gluons via heavy quark and squark loops (c,d).  Processes which contribute to the SD scattering include squark exchange (a) and $Z$ boson exchange (e).  However, SD predictions are far less constrained than SI.}
\label{fig:fd}
\end{center}
\end{figure}
\noindent Here $G^{p}_{s}$ is given in terms of the hadronic matrix elements $\langle N|\bar q q | N\rangle$ and couplings by \cite{ds}
\bea
G^{p}_{s}&=&\sum_{q=u,d,s,c,b,t}\langle N|\overline{q}q|N\rangle \frac{1}{2}\sum_{k=1}^{6}\frac{g_{\tilde{q_L}_{k}\chi q}g_{\tilde{q_R}_{k}\chi q}}{m^{2}_{\tilde{q}_{k}}}\\
&-&\sum_{q=u,d,s}\left(\langle N|\bar q q|N\rangle \sum_{h=H_{1},H_{2},H_{3}}\frac{g_{h\chi\chi}g_{hqq}}{m^{2}_{h}}\right)-{2\over27}\sum_{q=c,b,t}\left(f_{TG}^{(p)} {m_p\over m_q} \sum_{h=H_{1},H_{2},H_{3}}\frac{g_{h\chi\chi}g_{hqq}}{m^{2}_{h}}\right)\nn ,
\eea
where the three terms correspond to the diagrams in Fig. \ref{fig:fd}(a,b,c), respectively.  We incorporated the following updated hadronic matrix elements to the DarkSUSY code taken from \cite{Ellis:2000ds}
\bea
\langle N|\overline{q}q|N\rangle&=&f^{p}_{Tq}\frac{m_{p}}{m_{q}},\\ 
f^{p}_{Tu}=0.020\pm0.004,\quad f^{p}_{Td}&=&0.026\pm0.005,\quad f^{p}_{Ts}=0.118\pm0.062, \\
f^p_{TG}&=&0.84\nn
\eea
We include effects from the exchanges of the scalar quarks of all six generations and the Higgs exchange from light quarks and from gluons via heavy quark loops.  The diagrams are shown in Fig. \ref{fig:fd}.  The scalar quark contributions are suppressed by our choice of TeV squark masses.  The uncertainty in the SI scattering cross-section are large, of order 60\%, due to the above uncertainties in the hadronic matrix elements.

\begin{figure}[tb]
\begin{center}
\includegraphics[width=0.35\textwidth,angle=-90]{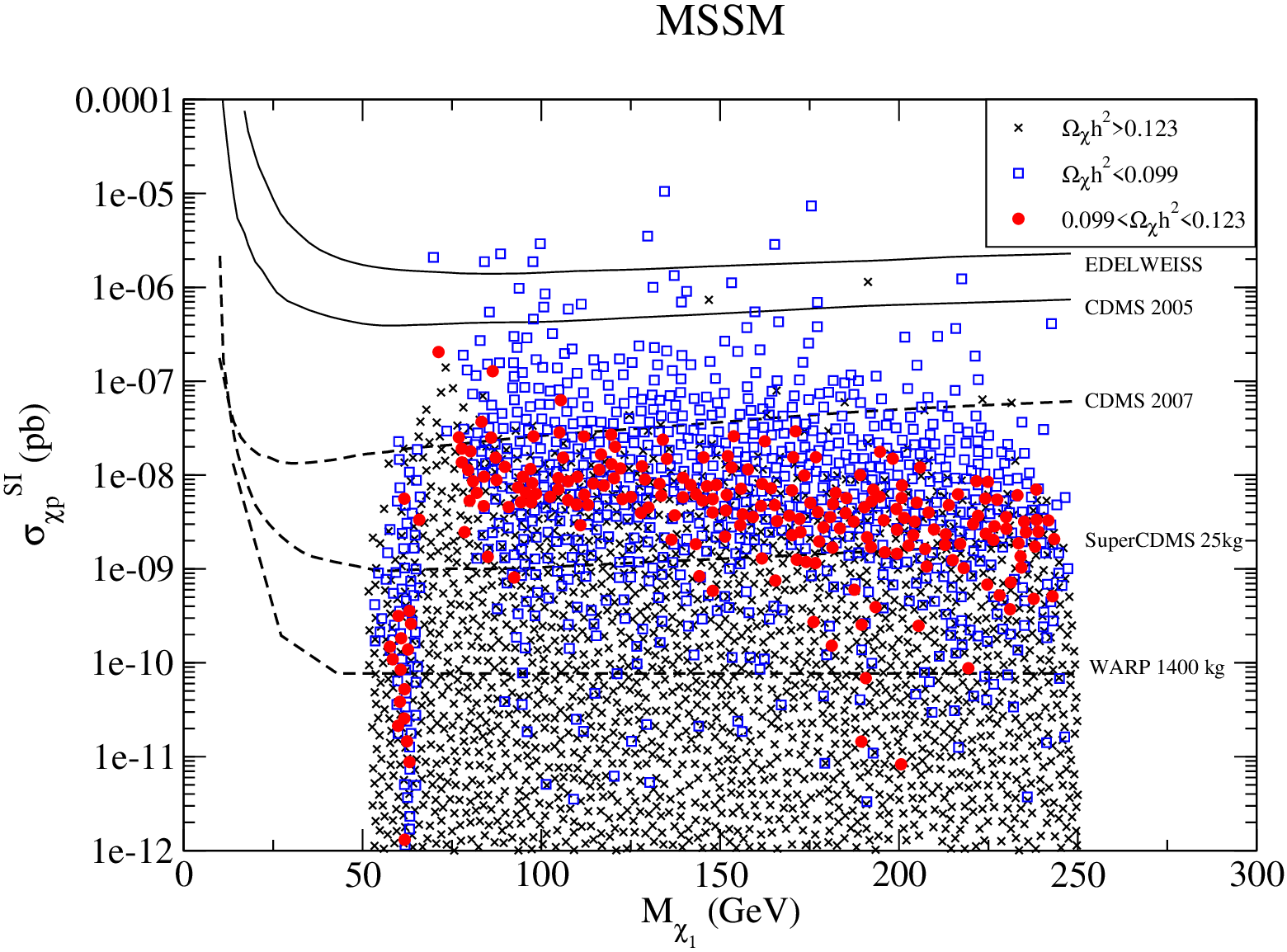}
\includegraphics[width=0.35\textwidth,angle=-90]{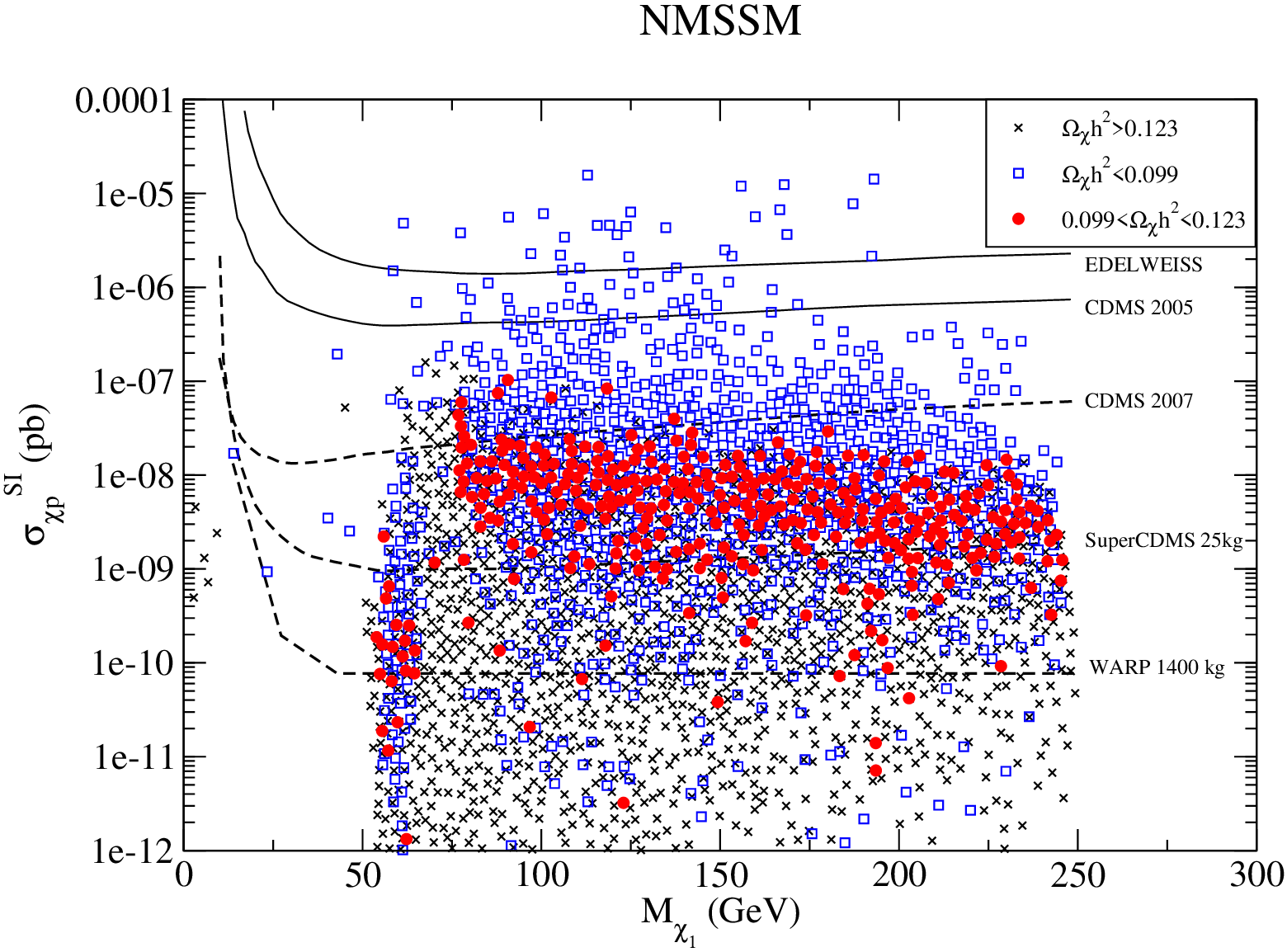}\\
(a)\hspace{0.48\textwidth}(b)
\includegraphics[width=0.35\textwidth,angle=-90]{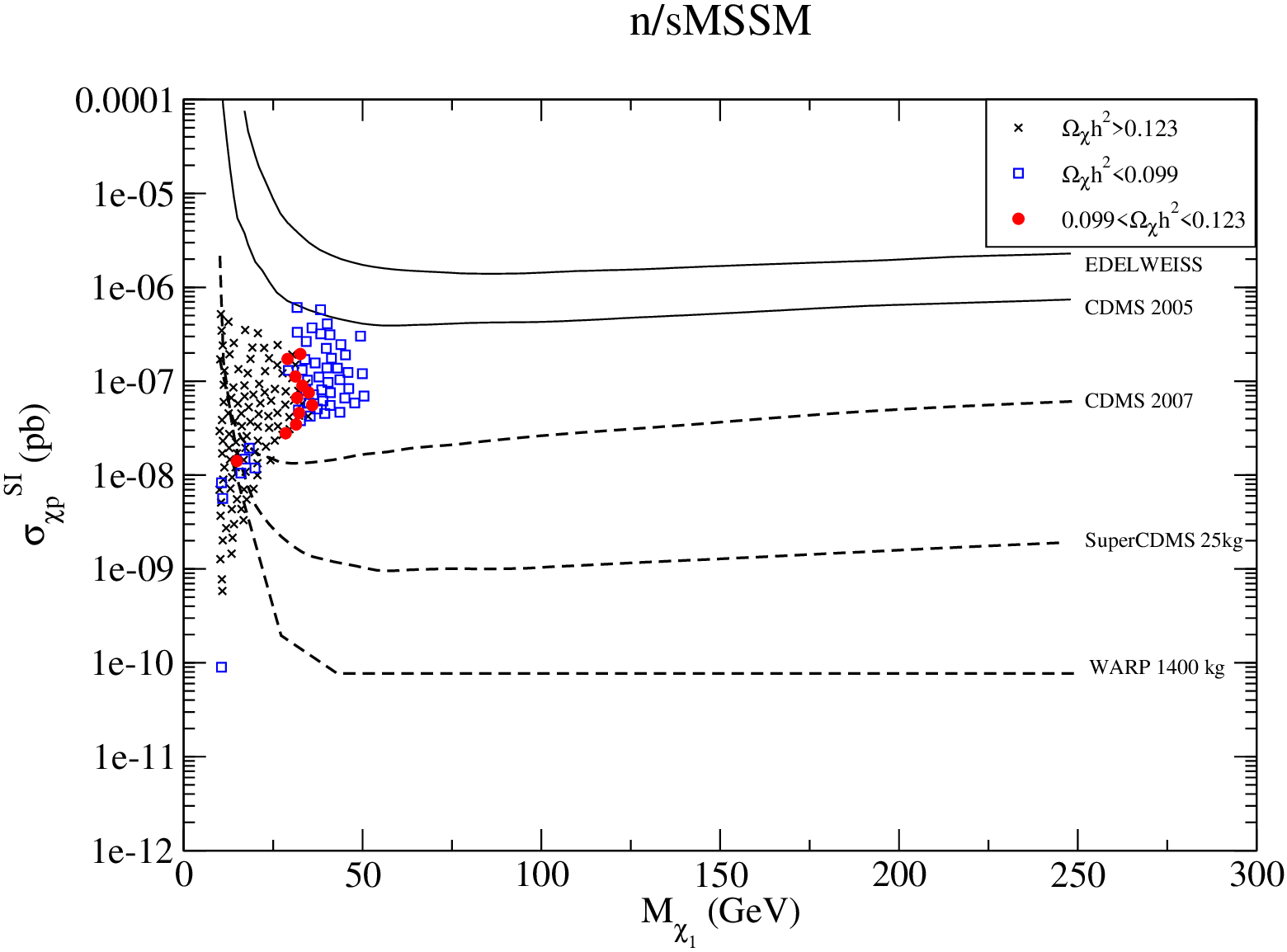}
\includegraphics[width=0.35\textwidth,angle=-90]{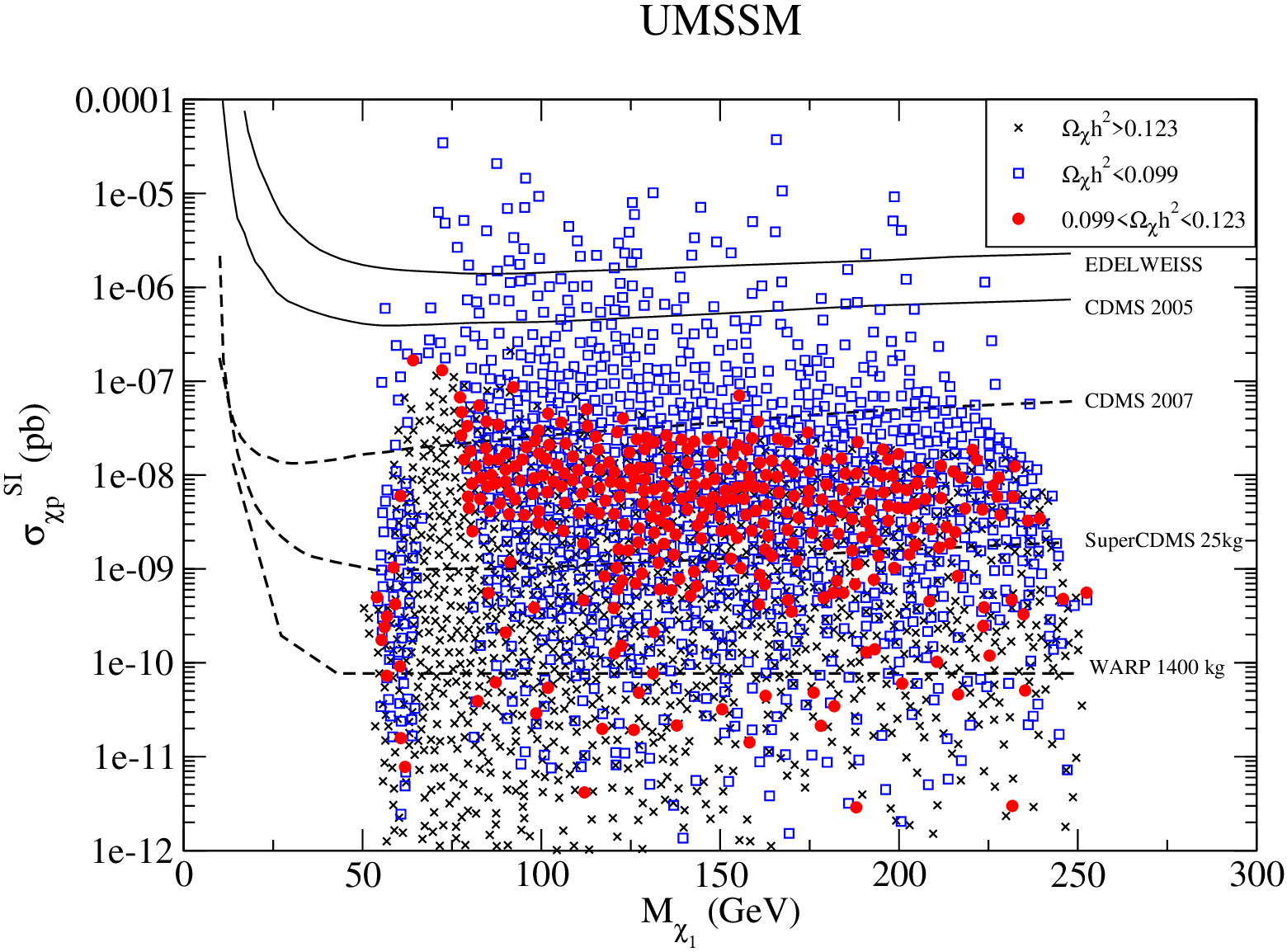}\\
(c)\hspace{0.48\textwidth}(d)
\caption{Expected SI direct detection cross-section for (a) MSSM, (b) NMSSM (c) n/sMSSM and (d) UMSSM.  The expected sensitivities of the EDELWEISS, CDMS II (2005), CDMS 2007, SuperCDMS (25 kg) and WARP (2.3L) experiments are shown.  Over most of the neutralino mass range, the experiments should detect the signals from the MSSM, NMSSM and UMSSM.  However, if the neutralino annihilates via a Higgs boson resonance, the relic density may be in the preferred region while the direct detection rate is out of reach of future experiments.}
\label{fig:scatt}
\end{center}
\end{figure}

In Fig. \ref{fig:scatt}, we show the predicted direct detection cross-section for the MSSM and the singlet extended models along with the sensitivities of EDELWEISS, CDMS II (2005), CDMS 2007, SuperCDMS (25 kg) and WARP (2.3L).   Over most of the region of neutralino mass, these experiments could find a signal for the MSSM, NMSSM and UMSSM.  There are more points where the lightest neutralino of the MSSM may be detectable via recoil experiments than typical CMSSM expectations since the general parameter treatment of the MSSM can yield scattering cross-section that are a few orders of magnitude larger than in the CMSSM \cite{Ellis:2005mb,Baer:2003jb,Trotta:2006ew}.

Many of the points consistent with the observed relic density are within the reach of SuperCDMS and WARP (2.3L).  The region below the WARP sensitivity corresponds to neutralino annihilation through the $H_1$ (or $H_2$) pole.  To account for the observed relic density, the $\N_1 \N_1 H_i$ coupling is small to balance the resonant enhancement of the annihilation rate.  This forces the scattering cross-section via the Higgs states to be small, and thus fall below future sensitivities.  

Since the n/sMSSM has a light singlino state, the strict limits on the allowed neutralino mass in this model from the expected relic density can be used to deduce a lower limit on the direct detection cross-section.  This is apparent in Fig. \ref{fig:scatt}(c) as most of the allowed region is above the expected sensitivity of CDMS 2007, which should allow CDMS to place extremely strong constraints on this model if no signal is found, assuming that the neutralino is responsible for a major part of the dark matter in the universe.  Still, detection could be precluded if annihilation through the Higgs resonance occurs with $m_{\N_1} < 30$ GeV and $M_{H_1} \simeq m_{\N_1} / 2$, as shown by the isolated low $\Omega_{\N_1}h^2$ points in Fig. \ref{fig:RELDEN}c, or by neutralino decays into an additional still lighter particle \cite{Feng:2004gg}.

The predictions of the SI cross-sections are heavily dependent on the local density of dark matter in our galaxy.  The present halo model has many assumptions and uncertainties.  Caustic rings are predicted to exist in the galaxy with a ring overlapping in the vicinity of our solar system \cite{Sikivie:2001fg}.  That could may increase the local dark matter velocity by a factor of three or more, resulting in a correspondingly enhanced flux and direct detection rate.

%%%%%%%%%%%%%%%%%%%%%%%%%%%%
\section{Conclusions}
\label{sect:concl}
%%%%%%%%%%%%%%%%%%%%%%%%%%%%

In our investigation of the relic density constraints and the direct detection capabilities of present and upcoming dark matter experiments, we modified the DarkSUSY code to include the additional couplings and processes of singlet extended MSSM models\footnote{Lighter squarks and sleptons would allow the possibility of additional coannihilation regions.}.  We also updated the DarkSUSY values for the hadronic matrix elements and experimental constraints.  Our analysis assumed multi-TeV masses of squarks and sleptons.  We generally found close similarities of the NMSSM and UMSSM predictions with those of the MSSM due to the similarity of the compositions of their light neutralino states.  However, the n/sMSSM showed exceptional differences due to very light neutralinos that are dominantly singlino.  Some notable results from out study include:

\bi
\item The observed relic density can be accounted for in all of the singlet extended models, mainly with mixed Higgsino-gaugino dark matter.  In some cases, the lightest neutralino is dominantly singlino, or a mixture of singlino and $Z'$-ino in the case of the UMSSM.  The predicted relic density can also match the observed value if the neutralinos annihilate through a Higgs boson resonance.  The annihilation via the lightest Higgs boson can enhance the rate sufficiently to yield the correct relic density, even though the neutralino is typically bino in this case.  

\item The neutralino in the n/sMSSM typically has a mass in the range $30 \textrm{ GeV} \lesssim m_{\chi^{0}_{1}} \lesssim 37 \textrm{ GeV}$ to account for the total relic density found by the WMAP collaboration.  The upper bound on $m_{\N_1}$ may extend to $50$ GeV if the neutralino relic density is below the observed DM value \cite{Barger:2004bz,Menon:2004wv,Barger:2005hb}.  Masses smaller than 30 GeV may be allowed if annihilation through the light CP-odd Higgs boson or neutralino decays occur to a still lighter state.  

\item The singlino/$Z'$-ino dominated neutralino in the UMSSM can account for the observed relic density.  Since the coupling between the lightest neutralino pair and the singlet dominated $H_2$ can be large and resonant enhancement of the annihilation cross-section via $H_2$ can occur, the relic density can fall into the observed $2\sigma$ range.
 
\item The MSSM, NMSSM, and UMSSM predict spin-independent proton scattering cross-sections that may be detectable at SuperCDMS and WARP and be consistent with the WMAP $\Omega_{DM}$ measurements.  However, the recoil predictions of some models may be undetectable by these experiments due to the small Higgs neutralino coupling.

\item The n/sMSSM SI proton scattering cross-sections are highly favored to be detectable at CDMS 2007 while being compatible with WMAP $\Omega_{DM}$ observations.  However, if neutralino annihilation occurs through a light Higgs, measurement of the scattering cross-section can fall below the sensitivities of future experiments.

\item Our MSSM predictions are more general than mSUGRA results.
\ei

\noindent In addition to these constraints from the relic density bounds our study of further constraints in the Appendix found

\bi
\item Perturbativity constraints on $\lambda$ from RGE evolution give upper bounds on $\lambda(M_{t})$ below unity and require $\tan\beta \gtrsim 1.9$.
\item Requiring that the xMSSM models are consistent with the anomalous magnetic moment of the muon, $\Delta a_{\mu}$, provides lower limits of $\tan\beta \gtrsim 5$, though there are still theoretical uncertainties.
\ei

%%%%%%%%%%%%%%%%%%%%%%%%%%%%
\section{Acknowledgments}
%%%%%%%%%%%%%%%%%%%%%%%%%%%%
We thank H. Baer for valuable conversations about the capabilities of future direct detection experiments.  This work was supported in part by the U.S.~Department of Energy under grant No. DE-FG02-95ER40896, by the Wisconsin Alumni Research Foundation, by the Friends of the IAS, and by the National Science Foundation grant No. PHY-0503584.
\appendix
%%%%%%%%%%%%%%%%%%%%%%%%%%%%
\section{Neutralino mass matrix}
\label{apx:neutmassmtx}
%%%%%%%%%%%%%%%%%%%%%%%%%%%%

The neutralino mass matrix of the singlet extended models is extended due to the additional singlino and $Z'$-ino states that mix with the MSSM gauginos and higgsinos.  In the $(\tilde B^0,\tilde W^0,\tilde H_d^0,\tilde H_u^0, \tilde S^0,\tilde Z'^0)$ basis, the mass matrix is
\be
\mathbf{M_{\chi^{0}}}=
\left( \begin{array}{cccccc}
M_{1} & 0 & -g_{1}v_{1}/2 & g_{1}v_{2}/2 & 0 & 0 \\
0 & M_{2} & g_{2}v_{1}/2 & -g_{2}v_{2}/2 & 0 & 0 \\
-g_{1}v_{2}/2 & g_{2}v_{2}/2 & 0 & -\mu_{\rm{eff}} & 
-\mu_{\rm{eff}}v_{2}/s & g_{Z'}Q'_{H_{1}}v_{1} \\
g_{1}v_{2}/2 & -g_{2}v_{2}/2 & -\mu_{\rm{eff}} & 0 & 
-\mu_{\rm{eff}}v_{1}/s & g_{Z'}Q'_{H_{2}}v_{2} \\
0 & 0 & -\mu_{\rm{eff}}v_{2}/s & -\mu_{\rm{eff}}v_{1}/s & 
\sqrt{2}\kappa s & g_{Z'}Q'_{S}s \\
0 & 0 & g_{Z'}Q'_{H_{1}}v_{1} & g_{Z'}Q'_{H_{2}}v_{2} & 
g_{Z'}Q'_{S}s & M_{1'} \\
\end{array} \right),
\label{eqn:neutmtx}
\ee
where the $M_i$ are the gaugino masses (assumed universal at the GUT scale), and $v_1,v_2$ and $s$ are respectively $\sqrt 2(\langle H_d\rangle,\langle H_u\rangle,\langle S\rangle)$.
%%%%%%%%%%%%%%%%%%%%%%%%%%%%
\section{\small{x}MSSM Couplings}
\label{apx:coup}
%%%%%%%%%%%%%%%%%%%%%%%%%%%%
Couplings of the singlet extended models are changed with respect to the MSSM due to the additional singlet and $Z'$ contributions.  

\bi 
\item The Yukawa the CP-even and CP-odd Higgs couplings are 
\be
g_{ddH_{i}}=g^{SM}_{ffh}\frac{R^{i1}_{+}}{\cos\beta}\quad
g_{uuH_{i}}=g^{SM}_{ffh}\frac{R^{i2}_{+}}{\sin\beta},
\ee
\be
g_{ddA_{i}}=i \gamma^5 g^{SM}_{ffh}\frac{R^{i1}_{-}}{\cos\beta}\qquad
g_{uuA_{i}}=i \gamma^5 g^{SM}_{ffh}\frac{R^{i2}_{-}}{\sin\beta},
\ee
where $R_+^{ij}$ and $R^{ij}_-$ are the rotation matries that diagonalize the CP-even and CP-odd Higgs mass-squared matrices, respectively, in the $(H_u,H_d,S)$ basis, and $g_{ffh}^{SM}$ are the corresponding couplings in the SM (see Ref. \cite{Barger:2006dh}).  
\ei

There are some couplings that do not have a standard model counterpart, as follows:
\begin{itemize}
\item The Higgs-Neutralino-Neutralino coupling constants are:
\bea
g_{H_{i}\chi^{0}_{1}\chi^{0}_{1}} & = & [(g_{1}N_{11}-g_{2}N_{12}-g_{1'}Q_{H_{d}}N_{16})N_{13} + \sqrt{2}\lambda N_{14}N_{15}]R^{i1}_{+}\nn \\
& + & [(g_{2}N_{12}-g_{1}N_{11}-g_{1'}Q_{H_{u}}N_{16})N_{14}+\sqrt{2}\lambda N_{13}N_{15}]R^{i2}_{+} \label{eqn:xxh}\\
& + & [-g_{1'}Q_{S}N_{16}N_{15}+\sqrt{2}\lambda N_{13}N_{14}-\sqrt{2}\kappa N_{15}N_{15}]R^{i3}_{+} \nn
\eea
\bea
g_{A_{i}\chi^{0}_{1}\chi^{0}_{1}} & = &i \gamma^5 [(g_{1}N_{11}-g_{2}N_{12}-g_{1'}Q_{H_{d}}N_{16})N_{13} + \sqrt{2}\lambda N_{14}N_{15}]R^{i1}_{-}\nn \\
& + & [(g_{2}N_{12}-g_{1}N_{11}-g_{1'}Q_{H_{u}}N_{16})N_{14}+\sqrt{2}\lambda N_{13}N_{15}]R^{i2}_{-} \\
& + & [-g_{1'}Q_{S}N_{16}N_{15}+\sqrt{2}\lambda N_{13}N_{14}-\sqrt{2}\kappa N_{15}N_{15}]R^{i3}_{-} \nonumber  ,
\eea
where $N_{ij}$ is the rotation matric that diagonalizes the neutralino mass matrix in Eq. \ref{eqn:neutmtx}.

\item The Left and Right Handed Neutralino-Quark-Squark coupling constants are:
\bea
g_{L\tilde{u}_k\chi^{0}u_{i}}&=&-\frac{Z^{ik*}_{\tilde{u}}}{\sqrt{2}}(\frac{1}{3}N_{11}g_{1}+N_{12}g_{2})-Y^{i}_{u}Z^{(i+3)k*}_{\tilde{u}}N_{14}\\
g_{R\tilde{u}_k\chi^{0}u_{i}}&=&\frac{2\sqrt{2}g_{2}}{3}Z^{(i+3)k*}_{\tilde{u}}N^{*}_{11}-Y^{i}_{u}Z^{ik*}_{\tilde{u}}N^{*}_{14}\\
g_{L\tilde{d}_k\chi^{0}d_{i}}&=&-\frac{Z^{ik}_{\tilde{d}}}{\sqrt{2}}(\frac{1}{3}N_{11}g_{1}-N_{12}g_{2})-Y^{i}_{d}Z^{(i+3)k}_{\tilde{d}}N_{13}\\
g_{R\tilde{d}_k\chi^{0}d_{i}}&=&\frac{-\sqrt{2}g_{2}}{3}Z^{(i+3)k}_{\tilde{d}}N^{*}_{11}-Y^{i}_{d}Z^{ik}_{\tilde{d}}N^{*}_{13}
\eea
where $Z^{ik}_{\tilde{q}}$ and $Z^{(i+3)k}_{\tilde{q}}$ are the rotation matrices for the left and right handed scalar quarks, respectively.   The Yukawa couplings, $Y^i_q$, are defined for quark $q_i$.

\item The trilinear Higgs coupling is modified due to the additional singlet state in
the nMSSM, NMSSM, and UMSSM.  The $H_iA_jA_k$ 
coupling can be found using the projection \cite{Barger:2006dh}
\be
C_{H_iA_jA_k}=P_{H_i}P_{A_j}P_{A_k}V,
\ee
where $V$ is the Higgs potential and the projection operators are
\bea
P_{H_j}=\frac{1}{\sqrt{2}}\bigg{(}R^{j1}_+\frac{\partial}{\partial \phi_d}+R^{j2}_+\frac{\partial}{\partial \phi_u}+R^{j3}_+\frac{\partial}{\partial \sigma}\bigg{)}\\
P_{A_k}=\frac{1}{\sqrt{2}}\bigg{(}R^{k1}_-\frac{\partial}{\partial \varphi_d}+R^{k2}_-\frac{\partial}{\partial \varphi_u}+R^{k3}_-\frac{\partial}{\partial \xi}\bigg{)}
\eea
In a similar manner the $H_iH_jH_k$ coupling is
\be 
C_{H_iH_jH_k}=P_{H_i}P_{H_j}P_{H_k}V,
\ee
and the $H_iH_jA_k$ coupling is
\be
C_{H_iH_jA_k}=P_{H_i}P_{H_j}P_{A_k}V.
\ee
\end{itemize}

%%%%%%%%%%%%%%%%%%%%%%%%%%%%
\section{Experimental Constraints}
\label{sect:const}
%%%%%%%%%%%%%%%%%%%%%%%%%%%%
We generated the Higgs and neutralino masses and couplings by scanning over the parameters given in Ref. \cite{Barger:2006dh} and diagonalizing the Higgs and neutralino mass matrices.  For each model, the following LEP experimental constraints are applied: The $ZZh$ coupling limits, the bound on the lightest chargino, the charged Higgs mass bound, the $A_h$ associated production search limits and the contribution to the invisible decay width of the $Z$ boson, and the limit on the $Z-Z'$ mixing in the UMSSM, as in Ref. \cite{Barger:2006dh}.  In addition, we considered limits from the anomalous magnetic moment of the muon, coupling perturbativity constraints and the improved limits on $\text{BF}(b\rightarrow s\gamma)$ \cite{Gomez:2006uv}.

%%%%%%%%%%%%%%%%%%%%%%%%%%%%
\subsection{Anomalous Magnetic Moment of the Muon}
%%%%%%%%%%%%%%%%%%%%%%%%%%%%
The experimentally measured value of the anomalous magnetic moment of the muon $a_{\mu} = (g - 2)_{\mu}$ shows a $3.4 \sigma$ deviation \cite{Hagiwara:2006jt,Davier:2007ua}
\be
\Delta a _{\mu} = a_{\mu}(\rm{exp}) - a_{\mu}(\rm{SM}) = (27.6 \pm 8.1) \times 10^{-10}
\ee
from the Standard Model prediction based on the use of $e^{+}e^{-} \to \pi \pi$ data from CMD-2 and KLOE.  The constraints on the UMSSM parameter space due to the $a_\mu$ deviation were previously studied in Ref. \cite{Barger:2004mr}.  We apply the above $a_{\mu}$ constraint here and find that the results do not vary significantly across models.  In particular, $\Delta a_{\mu}$ gives modest lower limits on $\tan\beta$ when using the parameter values in Ref. \cite{Barger:2006dh} in a scan over the soft slepton masses $300\text{ GeV} \le M_{\widetilde{L}} = M_{\widetilde{E}}\le 2\text{ TeV}$.  At the $2\sigma$ level the $\Delta a_{\mu}$ bound limits $\tan\beta \gtrsim 5$.  We did not impose this constraint in our analyses, because there are still theoretical uncertainties involving the disagreement between estimates using $e^+e^-$ and $\tau$ decay data \cite{Hagiwara:2006jt,Davier:2007ua}.

\begin{figure}[tb]
\centering
\includegraphics[width=0.6\textwidth,angle=-90]{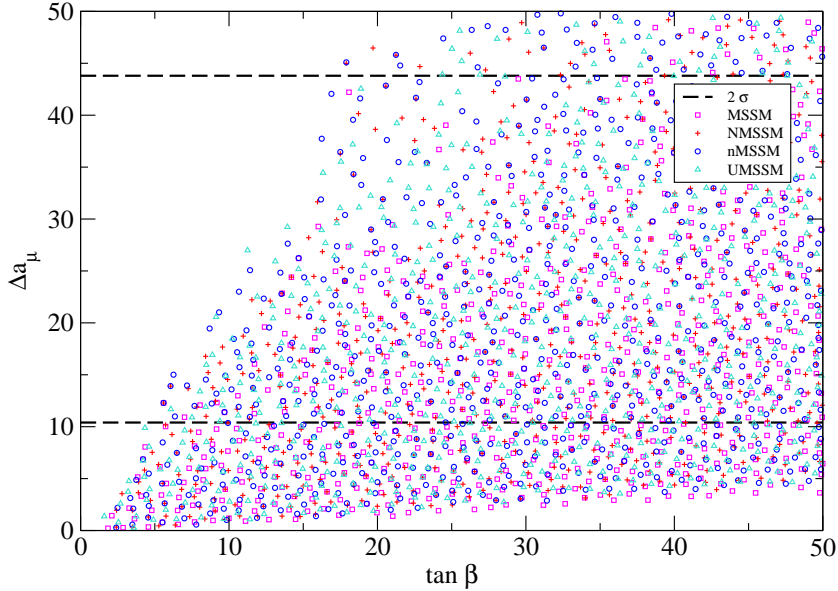}
\caption{The deviation of the anomalous magnetic moment of the muon $\Delta a_{\mu}$ consistent with the $2\sigma$ uncertainty for values of $\tan \beta$ in various models is within the dashed horizontal lines.
}
\label{fig:gm2}
\end{figure}

%%%%%%%%%%%%%%%%%%%%%%%%%%%%
\subsection{Perturbativity constraints}
%%%%%%%%%%%%%%%%%%%%%%%%%%%%
The singlet models that we consider may have  gauge coupling unification at a Grand Unification scale $M_{GUT} \sim 2 \times 10^{16}$ GeV.  We may require that the gauge and Yukawa couplings remain perturbative up to this scale.  To impose this constraint, we evaluate the running of the couplings in each model according to the renormalization group equations (RGEs) \cite{Ellis:1988er,King:2005jy,King:2005my,King:2007uj,Langacker:1998tc,Yeghian:1999kr,Barger:1992ac,Martin:1993zk,Ellwanger:1999ji,Ellwanger:2000fh,Elliott:1993bs,Binetruy:1991mk} at one loop order:
\bea
\frac{d g_{i}}{dt} & = & \beta_{i} g_{i}^{3}
,\nonumber\\
\frac{d g_{1'}}{dt} & = & \beta_{1'} g_{1'}^{3}
,\nonumber\\
\frac{d \kappa}{dt} & = & 6 \kappa \left[
\lambda^{2} + \kappa^{2} 
\right]
,\\
\frac{d \lambda}{dt} & = & \lambda \left[
4 \lambda^{2} + 2 \kappa^{2} + 3 h_{t}^{2} 
- 3 g_{2}^{2} - {3\over 5}g_{1}^{2} 
- 2 g_{1'}^{2} (Q_{S}^2 + Q_{H_d}^2  + Q_{H_u}^2)
\right]
,\nonumber\\
\frac{d h_{t}}{dt} & = & h_{t} \left[
  \lambda^{2} + 6 h_{t}^{2} 
- \frac{16}{3} g_{3}^{2} - 3 g_{2}^{2} - \frac{13}{15} g_{1}^{2}
- 2 g_{1'}^{2} (Q_{u}^2 + Q_{Q}^2  + Q_{H_u}^2)
\nonumber
\right]
\label{eqn:rges}
\eea
\noindent where $t=\frac{1}{(4\pi)^{2}}\ln(\mu / m_{t})$.  Here $h_{t}$ is the top quark Yukawa coupling and $Q_{H_u},Q_{H_d},Q_{S},Q_{u},$ and $Q_{Q}$ are the $U(1)'$ charges of the up and down Higgs doublets, the Higgs singlet, the up type quark and the quark doublet, respectively.  For specific models, the parameters $g'_1$ and $\kappa$ are appropriately turned off.

The scale factors and $\beta$-functions are given in Table \ref{tab:scaleAndBeta}.\footnote{We do not include the effect of kinetic mixing in the UMSSM  \cite{Babu:1996vt}.}.  In Table \ref{tab:charges} we give the quantum numbers under the Standard Model gauge groups, $U(1)_{\chi}$ and $U(1)_{\psi}$ (symmetries arising when the $E_{6}$ model is broken \cite{Langacker:1998tc}), and of $U(1)'$ (the additional symmetry of interest in the UMSSM).  As $U(1)'$ is simply a linear combination of $U(1)_{\chi}$ and $U(1)_{\psi}$, the quantum numbers under this symmetry are given by
\be
Q = Q_{\chi} \cos \theta_{E6} + Q_{\psi} \sin \theta_{E6}
\ee
\noindent and thus depend on the particular value of $\theta_{E6}$.

The RGEs were evolved from the low scale, chosen to be $m_{t}=171$ GeV, up to $M_{GUT} \sim 2\times 10^{16}$ GeV. The gauge couplings were first run independently up to $m_{t}$ from $M_{Z}$, using \cite{Yao:2006px,Bethke:2006ac}
\bea
\sin^{2} \theta_{W} (M_{Z}) &=& 0.23122 \pm 0.00015 \nonumber\\
\alpha(M_{Z}) &=& 1/(127.918 \pm 0.018) \\
\alpha_{S}(M_{Z}) &=& 0.1189 \pm 0.0010  \nonumber
%\alpha_{S}(M_{Z}) &=& 0.1216 \pm 0.0017  \nonumber
\eea
where $\alpha = e^{2} / 4\pi$, $\alpha_{S} = g_{3}^{2} / 4\pi$, $g_{1} = \sqrt{5\over 3} g_Y = \sqrt{5\over 3} e/ \cos \theta_{W}$, and $g_{2} = e / \sin \theta_{W}$.  The  gauge couplings were assumed to unify at the GUT scale, with the canonical normalization $g_{1'} =  g_{1}=g_2=g_3$ at the unification scale\footnote{$\beta_{1'}$ depends weakly on $\theta_{E_6}$, which is scanned, because of the $\hat H_3, \hat{\bar H_3}$ pair needed for gauge unification \cite{Langacker:1998tc}.  This implies that ${g}_{1'}=\sqrt{\lambda_{g}} g_1$ at the weak scale, where $\lambda_g$, which depends on $\theta_{E_6}$, is typically within 6\% of unity.}.  From the top quark pole mass, we obtained the low-scale value of the top quark Yukawa coupling
\be
h_{t}(m_{t}) = \frac{m_{t}(m_{t})}{\frac{1}{\sqrt{2}} v \sin \beta}
\ee

For the couplings to remain perturbative up to $M_{GUT}$, we require
\be
\kappa^{2} \le 4\pi \quad ; \quad
\lambda^{2} \le 4\pi \quad; \quad
h_{t}^{2} \le 4\pi
\label{eq:pertlim}
\ee

\begin{figure}[tb]
\centering
\includegraphics[width=0.6\textwidth,angle=-90]{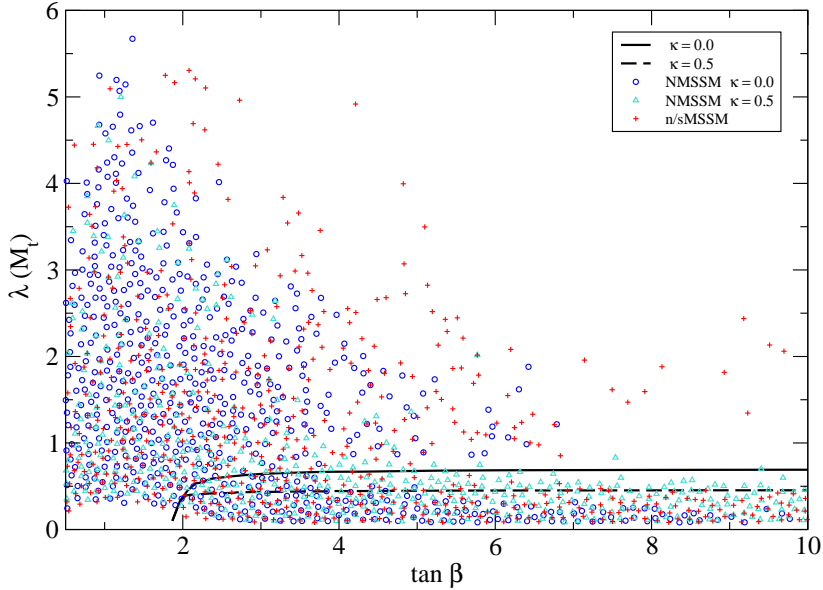}
\caption{Values of $\lambda$ vs. $\tan \beta$ consistent with all considered constraints except perturbativity in scans over all parameters.  The solid and dashed lines represent the maximum allowed values of $\lambda(m_{t})$ that remains perturbative in the NMSSM for $\kappa=0$ and  $\kappa=0.5$.  The n/sMSSM constraint is given by the $\kappa = 0$ result of the NMSSM.}
\label{fig:pertN}
\end{figure}
\begin{figure}[tb]
\centering
\includegraphics[width=0.6\textwidth,angle=-90]{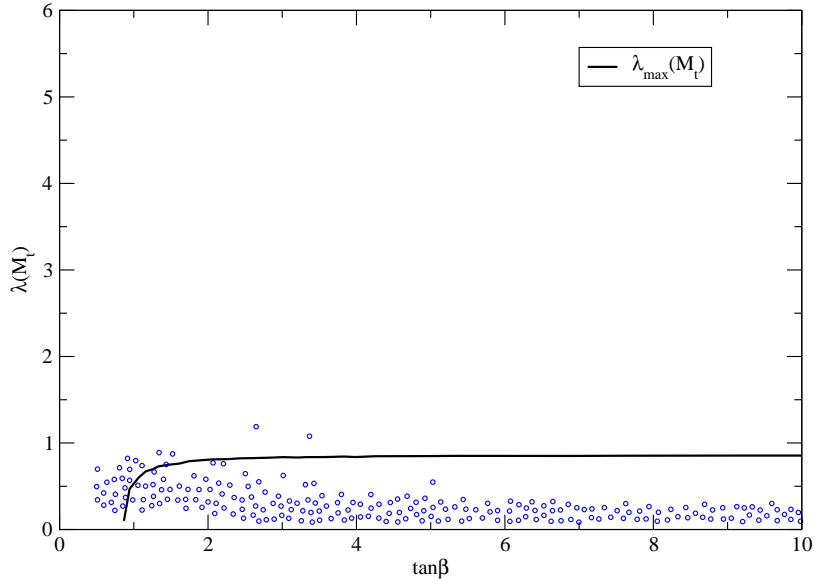}
\caption{Values of $\lambda$ vs. $\tan \beta$ consistent with all considered constraints except perturbativity in scans over all parameters in the UMSSM, including $\theta_{E6}$.  The solid line represents the maximum allowed values of $\lambda(m_{t})$ that remains perturbative for the particular choice of $\theta_{E6} = \arctan{\sqrt{15}}$.  Unlike the other models, $\lambda$ is strongly constrained by the requirement of a small $Z-Z'$ mixing~\cite{Barger:2006dh}, which, for reasonable $\mu_{\rm eff}$, favors large $\langle S \rangle$ and small $\lambda$. This is especially true for large $\tan \beta$ or small $\theta_{E_6}$, for which there are no cancellations between the contributions of $H^0_d$ and $H^0_u$ to the mixing.}
\label{fig:pertU}
\end{figure}
\noindent The NMSSM perturbativity constraints are shown in Figure \ref{fig:pertN} as the solid line for $\kappa = 0$ and a dashed line for $\kappa = 0.5$.  The n/sMSSM corresponds to the $\kappa =0$ limit of the NMSSM.  There are many points with large values of $\lambda$ at low $\tan\beta$ that violate the limit in the $\kappa = 0$ NMSSM and n/sMSSM.   Figure \ref{fig:pertN} also demonstrates that perturbativity becomes slightly more constraining in the NMSSM for larger $\kappa$, as $\kappa$ contributes to the running of $\lambda$.  The maximum allowed values of $\lambda(m_{t})$ for moderate values of $\tan\beta$ are given in Table \ref{tab:lambdaNMSSM} for several values of $\kappa$.  These values are slightly lower than the constraints of $\lambda(m_{t}) \lesssim 0.75$ and $\sqrt{\lambda^{2}(m_{t}) + \kappa^{2}} \lesssim 0.75$ for the NMSSM that are often used as approximations to the RGE running \cite{Barger:2006dh}\footnote{Note that these constraints can be weakened if one adds additional vector-like matter, increasing the (negative) effect of the gauge couplings in \ref{eqn:rges} \cite{King:2005jy,King:2005my,King:2007uj,Langacker-kang}.}.

The perturbativity limit for the UMSSM is given in Figure \ref{fig:pertU} for the example\footnote{The value for which the $U(1)'$ charge of the right handed neutrino $N_i^c$ vanishes.} of $\theta_{E6} = \arctan{\sqrt{15}}$.  It is less constraining than the NMSSM, with very few models being eliminated due to $\lambda_{max}(m_{t})$, which varies from $\sim 0.82$ to $\sim 0.86$, depending on the choice of $\theta_{E6}$.

\begin{table}[tb]
\caption{The maximum allowable value of $\lambda$ for perturbativity in the NMSSM, as a function of $\kappa$.}
\begin{center}
\begin{tabular}{|c|c|c|c|c|c|}
\hline \hline
$ \kappa $ & $0.0$ & $0.3$ & $0.4$ & $0.5$ & $0.6$ \\ \hline 
$\lambda_{max}$ & $0.70$ & $0.64$ & $0.58$ & $0.47$ & $0.25$ \\ \hline
\hline
\end{tabular}
\end{center}
\label{tab:lambdaNMSSM}
\end{table}%

In addition to the perturbativity upper bound on $\lambda(m_{t})$, the RGE running sets a lower limit of $\tan\beta \gtrsim 1.8 - 1.9$ for the NMSSM/nMSSM and $\tan\beta \gtrsim 0.8 - 0.9$ for the UMSSM.  Below these $\tan \beta$ values, a large $h_t$ often feeds into the running of $\lambda$, making $\lambda$ non-perturbative and sometimes resulting in a Landau pole below $M_{GUT}$ \cite{Ellwanger:1999ji,Ellwanger:2000fh,Elliott:1993bs,Yeghian:1999kr}.

\begin{table}[tb]
\caption{Matter multiplets in the UMSSM and their various quantum numbers (according to \cite{Langacker:1998tc}, but noting our difference in normalization).  The index $i$ sums over generations.}
\begin{center}
\begin{tabular}{|c|c|c|c||c|c|c|}
\hline \hline
superfields & $SU(3)_{c}$ & $SU(2)_{L}$
& $Q_Y=\sqrt{5\over 3} Q_1$ & $U(1)'$
& $\sqrt{40}Q_\chi$ & $\sqrt{24}Q_\psi$
 \\ \hline
$\hat{Q}_{i}$       & $\bf{3}$ & $\bf{2}$ & $\frac{1}{6}$  & $Q_{Q}$        
& $-1$ & $1$ \\ \hline
$\hat{u}_{i}^{c}$   & $\bf{3}$ & $\bf{1}$ & $-\frac{2}{3}$ & $Q_{u}$        
& $-1$ & $1$ \\ \hline
$\hat{d}_{i}^{c}$   & $\bf{3}$ & $\bf{1}$ & $\frac{1}{3}$  & $Q_{d}$        
& $3$  & $1$ \\ \hline
$\hat{L}_{i}$       & $\bf{1}$ & $\bf{2}$ & $-\frac{1}{2}$ & $Q_{L}$        
& $3$  & $1$ \\ \hline
$\hat{E}_{i}^{c}$   & $\bf{1}$ & $\bf{1}$ & $1$            & $Q_{E}$        
& $-1$ & $1$ \\ \hline
$\hat{H}_{1i}$      & $\bf{1}$ & $\bf{2}$ & $-\frac{1}{2}$ & $Q_{H_{d}}$
& $-2$ & $-2$\\ \hline
$\hat{H}_{2i}$      & $\bf{1}$ & $\bf{2}$ & $\frac{1}{2}$  & $Q_{H_{u}}$
& $2$  & $-2$\\ \hline
$\hat{D}_{i}$       & $\bf{3}$ & $\bf{1}$ & $-\frac{1}{3}$ & $Q_{D}$        
& $2$  & $-2$\\ \hline
$\hat{\bar{D}}_{i}$ & $\bf{3}$ & $\bf{1}$ & $\frac{1}{3}$  & $Q_{\bar{D}}$  
& $-2$ & $-2$\\ \hline
$\hat{S}_{i}$       & $\bf{1}$ & $\bf{1}$ & $0$            & $Q_{S}$        
& $0$  & $4$ \\ \hline
$\hat{N}_{i}^{c}$   & $\bf{1}$ & $\bf{1}$ & $0$            & $Q_{N}$        
& $-5$ & $1$ \\ \hline
$\hat{H}_{3}$       & $\bf{1}$ & $\bf{2}$ & $-\frac{1}{2}$ & $-Q_{H_{u}}$    
& $-2$ & $2$ \\ \hline
$\hat{\bar{H}}_{3}$ & $\bf{1}$ & $\bf{2}$ & $ \frac{1}{2}$ & $ Q_{H_{u}}$
& $2$  & $-2$ \\ \hline 
\hline
\end{tabular}
\end{center}
\label{tab:charges}
\end{table}%

\begin{table}[tb]
\caption{One-loop $\beta$ functions for the three models \cite{Langacker:1998tc,Yeghian:1999kr}.  The traces involve sums over all matter superfields and generations.}
\begin{center}
\begin{tabular}{|c|c|c|c|c|c|}
\hline \hline
&  $\beta_{1}$ & $\beta_{2}$ & $\beta_{3}$ & $\beta_{1'}$ \\ \hline 
NMSSM/nMSSM &  $33/5$ & $1$ & $-3$ & $0$ \\ \hline
UMSSM       &  ${3\over 5} Tr Q_Y^2 = 9+{3\over 5}$
& $4$ & $0$  & $TrQ^{2}=9+4 Q_{H_u}^2$ \\ \hline
\hline
\end{tabular}
\end{center}
\label{tab:scaleAndBeta}
\end{table}%

\newpage
%%%%%%%%%%%%%%%
\bibliographystyle{h-physrev}
%\newpage
\bibliography{ds-xmssm}                      %name of .bib file
%%%%%%%%%%%%%%%s
\newpage

\end{document}